\documentclass[iop]{emulateapj}
\usepackage{natbib} 
\usepackage{xcolor}

\slugcomment{Draft for the Astrophysical Journal}

\shorttitle{Secondary Eclipse of CoRoT-2b} \shortauthors{Wilkins et al.}

\begin{document}

\title{The Emergent 1.1-1.7 Micron Spectrum of the Exoplanet CoRoT-2b as
Measured Using the Hubble Space Telescope}

\author{Ashlee~N.~Wilkins\altaffilmark{1}, Drake~Deming\altaffilmark{1},
Nikku~Madhusudhan\altaffilmark{2}, Adam~Burrows\altaffilmark{3},
Heather~Knutson\altaffilmark{4}, Peter~McCullough\altaffilmark{5} \&
Sukrit~Ranjan\altaffilmark{6}}

\altaffiltext{1}{Department of Astronomy, University of Maryland, College Park, MD 20742 (USA); awilkins@astro.umd.edu}
\altaffiltext{2}{Institute of Astronomy, University of Cambridge, Cambridge CB3 0HA (GB)}
\altaffiltext{3}{Department of Astrophysical Sciences, Princeton University, Princeton, NJ 08544-1001 (USA)} 
\altaffiltext{4}{Division of Geological and Planetary Sciences, California Institute of 
  Technology, Pasadena, CA 91125 (USA)}
\altaffiltext{5}{Space Telescope Science Institute, Baltimore MD 21218 (USA)}
\altaffiltext{6}{Harvard-Smithsonian Center for Astrophysics, Cambridge, MA 02138 (USA)}

\begin{abstract} We have used Hubble/WFC3 and the G141 grism to
  measure the secondary eclipse of the transiting very hot Jupiter
  CoRoT-2b in the 1.1-1.7\,$\mu$m spectral region.  We find an eclipse
  depth averaged over this band equal to $395^{+69}_{-45}$\,parts per
  million, equivalent to a blackbody temperature of $1788\pm18$K.  We
  study and characterize several WFC3 instrumental effects, especially
  the ``hook'' phenomenon described by \citet{deming13}.  We use data
  from several transiting exoplanet systems to find a quantitative
  relation between the amplitude of the hook and the exposure level of
  a given pixel.  Although the uncertainties in this relation are too
  large to allow us to develop an empirical correction for our data,
  our study provides a useful guide for optimizing exposure levels in
  future WFC3 observations.  We derive the planet's spectrum using a
  differential method.  The planet-to-star contrast increases to
  longer wavelength within the WFC3 bandpass, but without water
  absorption or emission to a $3\sigma$ limit of 85 ppm. The slope of the WFC3 spectrum is significantly less than the slope of the  best-fit blackbody.  We compare all existing eclipse data for this
  planet to a blackbody spectrum, and to spectra from both solar
  abundance and carbon-rich (C/O=1) models.  A blackbody spectrum is
  an acceptable fit to the full dataset.  Extra continuous opacity due to
  clouds or haze, and flattened temperature profiles, are strong
  candidates to produce quasi-blackbody spectra, and to account for
  the amplitude of the optical eclipses.  Our results show ambiguous
  evidence for a temperature inversion in this planet.
\end{abstract}

\keywords{stars: individual (CoRoT-2) -- planets and satellites: individual (CoRoT-2b) -- planets and satellites: atmospheres -- techniques: photometric -- techniques: spectroscopic}
\section{Introduction}

Very Hot Jupiters are gas-giant exoplanets with orbital periods less than
about 3 days.  The close proximity of VHJs to their host stars enhances the
influence of irradiation, tidal forces, and stellar activity on their structure
and evolution. CoRoT-2b \citep{corot} is a VHJ of particular interest because of
lingering questions about the structure of its atmosphere, which can be studied
with observations of its secondary eclipse.  \citet{alonso09} announced the
first secondary eclipse observations of CoRoT-2 in the~\emph{CoRoT} optical
waveband, followed by the mid-infrared~\emph{Spitzer} secondary eclipse
measurements of \citet{gillon10}, re-analyzed and expanded with
Warm~\emph{Spitzer} eclipses by \citet{drakespitzer}. \citet{alonso10} added a
secondary eclipse point in the K$_{s}$ band. The analysis of \citet{gillon10}
favored a poor day-night-side heat distribution in CoRoT-2b's atmosphere. 
\citet{drakespitzer} found a high 4.5\,$\mu$m flux as the only disagreement with
a solar-composition, equilibrium chemistry model of the atmospheric temperature structure.
\citet{drakespitzer} considered possible emission in the 4.5\,$\mu$m band from
CO mass loss. Both works question, but do not rule out, the presence of a
temperature inversion in the atmosphere caused by an upper atmosphere absorber.
\citet{madhu12} finds that either a carbon-rich or solar abundance non-inverted
model fits the data available in the literature.

These widely varied, competing explanations for this planet
demonstrate the importance of spectroscopic observations. CoRoT-2b
clearly does not fit the standard solar-composition, equilibrium
chemistry model that satisfactorily describes many planets in its
class, and we explore the anomalous spectral shape. For a clear
illustration of CoRoT-2b's standing as an outlier among VHJs, see
\citet{knutson10}. CoRoT-2 is a very active star, a young Solar analog, and yet a temperature inversion cannot be ruled out and the planet does not fit clearly into the otherwise well-defined inverted/non-inverted planet classifications. This curious state of the planet is perhaps due to a magnetic interaction between the planet \citep{lanza09} and CoRoT-2. Any further understanding would require more measurements of the planet in new wave bands.

\begin{deluxetable*}{cccc}[t!]
\centering 
\tablecaption{CoRoT-2 Observation Summary} 
\tablehead{\colhead{Visit} & \colhead{UT Date \& Time
(hr:min-hr:min)} & \colhead{Number of Exposures} & \colhead{Orientation Angle}}
\startdata 
A & 10-18-2010 11:12-16:45 & 271 & 80.4$^{\circ}$\\ 
B & 9-16-2011 09:37-15:07 & 276 & 93.9$^{\circ}$\\ 
C & 9-23-2011 07:41-13:11 & 275 & 90.7$^{\circ}$
\enddata 
\label{tab:corot2obs} 
\end{deluxetable*} 

In this paper, we use the G141 infrared grism on the Hubble Space
Telescope's Wide-Field Camera~3 (HST's WFC3) to detect the day-side thermal
emission spectrum of CoRoT-2b from 1.1\,$\mu$m to 1.7\,$\mu$m. The
CoRoT-2 system is part of an HST Cycle-18 program that observed a wide
range of HJs/VHJs in transit and secondary eclipse, and gives us the
basis for new insights into the instrumental effects of WFC3
\citep{deming13, huitson, line, mandell, ranjan13}. In what follows, we
  describe the observations of the CoRoT-2 system in
  \S\ref{sec:observations} and the initial stages of data analysis in
  \S\ref{sec:red}. In \S\ref{sec:systematics} we place our
  observations in the larger context of other HST programs with WFC3
  in order to provide a comprehensive systematic description of the
  instrumental effects encountered in these observations. We then
  present our methods of obtaining the band-integrated secondary
  eclipse curve (\S\ref{sec:whitelight}) and derivation of the
  spectrum (\S\ref{sec:spectrum}) of CoRoT-2b. Finally, we use our
  results to constrain models for the atmosphere of the planet in
  \S\ref{sec:models}, and we summarize in \S{8}.

\section{Observations}\label{sec:observations} 
We observed CoRoT-2 using the G141 grism of WFC3 (1.1-1.7\,$\mu$m), in
three separate visits, each comprising four orbits of HST and hereafter called visits A, B, and C.  We used the 128$\times$128-pixel subarray of the 1024$\times$1024-pixel
detector. At the beginning of each visit, we acquired a single direct
image of the system with the F139M filter, a medium-band filter
centered at 1.39\,$\mu$m; the location of the target in this direct
image defines the initial wavelength solution for the grism spectra.
A summary of the observations is in Table~\ref{tab:corot2obs}.

Most of our observations in program 12181, including those of CoRoT-2, were executed before the
advent of spatial scan mode \citep{pm2}.  Lacking the spatial scan, WFC3
observations of relatively bright stars can be inefficient, because
the time required to transfer the data greatly exceeds the exposure
time for bright exoplanet host stars.  We maximize the efficiency by
using subarrays and by exposing the detector to fluence levels
approaching or equaling saturation.  Even at a saturated exposure
level, an unsaturated signal is available because the detector is
sampled `up the ramp' multiple times within each exposure, and all the
samples are saved in the data. Isolating less than the full number of
samples, a linear signal can be obtained even in the saturated
case. Our CoRoT-2 grism data are exposed so that the brightest pixel
contains about 70,000 electrons in a full exposure, which is
approximately the level of 5\% non-linearity.

\section{Initial Data Analysis}\label{sec:red}

In order to explore whether our results are sensitive to details of
the data analysis, we use two parallel but independent methods to
process the data. To avoid confusion with the visit terminology (A, B,
C), we denote the two methods as $\alpha$ and $\beta$.  Method
$\alpha$ makes more explicit corrections and manipulations of the data
than does method $\beta$.  Exoplanet signals are subtle, and the more
the data are processed, the more the potential for adding numerical
noise that may mask the small exoplanet signal or even fooling oneself
into detecting a false signal.  Our dual-track analysis allows us to
evaluate the trade-off between the most `complete' method versus the
potential for degrading the results by over-processing of the data.
It also allows us to evaluate what corrections are necessary, and what
corrections can be neglected. Upon measurement of the eclipse curve,
the methods yield consistent results.

Method $\alpha$ uses ``flt" FITS image files retrieved from the Mikulski Archive
at Space Telescope (MAST) server, located at the Space Telescope Science
Institute (STScI). The ``flt" files were calibrated through the WFC3 pipeline's
high-level task, calwf3, which includes two low-level tasks, wf3ir and wf3rej,
that apply to the infrared channel.  wf3ir performs standard calibrations,
including corrections for bias, non-linearity, dark current, and bad pixels due
to energetic particle hits, while wf3rej completes more bad pixel rejection and
combines images. \citet{dhb} gives details of this pipeline.  We multiply the
resultant signal rates (electrons per second) by the integration time to infer
the accumulated signal on each pixel, in electrons.

Method $\beta$ begins with ``ima" FITS files from the MAST
server. These files give the `sample-up-the ramp' values of each pixel
at 4 times during each 22-second exposure, and are processed to
correct for non-linearity, but not to reject energetic particle hits.
We process these files (minimally) by fitting a linear slope to the
four samples as a function of time for each pixel, to determine the
rate at which electrons are accumulating in the pixel.  Our linear fit
weights each sample of a given pixel by the square-root of the signal
level, as appropriate for Poisson errors.  Multiplying the fitted
slope by the 22-second integration time yields the accumulated signal
in electrons. This process does not include any correction for
energetic particle hits. Rather, we correct those at later stages
of the $\beta$-analysis, and we also evaluate the success of the
non-linearity corrections by repeating the $\beta$-analysis and
restricting the linear fit versus sample time to only the first three
samples.

Using the smaller subarray means the grism data consists of the central 128 pixel columns of the first-order spectrum out of the 150 on a larger (sub)array. Nevertheless, using the 128 subarray increased the efficiency of the observations (i.e., minimizing data transfer time on the spacecraft), more than
justifying the loss of points at the edges of the grism response.

To extract the spectrum of the star+planet system, we sum the pixels after
background subtraction, using a box defining a range in rows. We adopt a box
size of height 61 pixels (a central pixel, plus 30 above and below it). The box
length is the full 128 pixel length of the subarray, but we later trim the
spectrum in wavelength.  We sum the box over rows to produce spectra, and we
further sum over wavelength to produce a `white light' photometric time series. 
The spectra are very stable in position (jitter less than several hundredths of
a pixel), and the intensity level falls by 2.5 orders of magnitude over the
30-pixel half-height of the box. Therefore we use fixed integral coordinates for
the box in each visit, and we weight each pixel equally when performing the sum.
 This spectral extraction is the same for both the $\alpha$ and $\beta$
analyses.

In the following, we discuss the various sub-elements of the data analysis
(\S\ref{sec:bp_cr},\ref{sec:bg}) including the wavelength calibration
(\S\ref{sec:wave}) and flat-fielding (flux calibration, \S\ref{sec:flux}), while
the more extensive task of characterizing the instrumental systematics is
discussed in \S\ref{sec:systematics}.

\subsection{Bad Pixel Correction}\label{sec:bp_cr} 

Bad pixel correction due to energetic particle hits is part of the
calwf3 processing used for our $\alpha$ analysis.  Additional pixels
not identified by calwf3 may still be erroneously high or low in value
and need correction. For both $\alpha$ and $\beta$ analyses, we
identify and correct bad pixels immediately prior to the spectral
extraction (i.e., before summing the box). Our $\alpha$ analysis
inspects pixels in each column of the spectral box (i.e., a single
wavelength) that deviate significantly from a Gaussian profile of the
spectral trace. Such deviations are virtually always characterized by
much higher intensity levels. Those pixels that are more than 10 times
greater than the fitted Gaussian value are replaced by a 7-pixel
median in the vertical direction (perpendicular to the dispersion) at
that wavelength.

Our $\beta$ analysis must be more sophisticated as regards bad pixels, since
these data have not been processed by calwf3.  We examine the ratio of a given
pixel to the total of all pixels in that row, i.e., the ratio of a single pixel
to the sum over wavelength at each spatial position.  Because of spatial
pointing jitter, pixel intensities can vary with time in an absolute sense, but
their relative variation should be similar at all wavelengths.  We examine the
ratio as a function of time (i.e., for each exposure) and we identify instances
where a given pixel does not scale with its row sum.  We identify $>4\sigma$
outliers, and correct them using a 5-frame median value of the ratio at that
time.

\subsection{Background Subtraction}\label{sec:bg} 

For both the $\alpha$ and $\beta$ analyses, we calculate the
background individually for each exposure by using pixels outside of
the spectral box. Specifically, the pixels used are those that lie
directly below the spectrum on the subarray, which is the section of
the subarray corresponding to the width of the spectrum and extending
from the bottom edge of the spectral box to the bottom edge of the
subarray. We construct a histogram of intensity values in these
pixels and fit a Gaussian to the histogram. The adopted background
value is the intensity corresponding to the central value of the
fitted Gaussian, and we do assume that it is independent of wavelength
to the limit of our precision. This is typically a few tens of
electrons per pixel, several orders of magnitude less than the signal
in the stellar spectrum, and the sum is thus also significantly lower
when calculating the white light curve and its corresponding
background.  Background subtraction therefore has a relatively minor
effect on our analysis.

\subsection{Wavelength Calibration}\label{sec:wave} 

Wavelength calibration utilizes both the direct image and the spectral
image, as the wavelength of a given pixel depends upon its location on
the detector relative to the direct image. \citet{cal} outline the
procedure for wavelength calibration in an STScI calibration
report. The equations governing the wavelength for a pixel at a given
x-position in the first-order spectrum are:

\begin{equation} \lambda (x) =
dldp_{0} + dldp_{1}\Delta x 
\end{equation} where 
\begin{displaymath} dldp_{0} = a^{0}_{0} + a^{1}_{0}x_{center} 
\end{displaymath} 
\begin{displaymath} dldp_{1} =
a^{0}_{1} + a^{1}_{1}x_{center} + a^{2}_{1}y_{center} + a^{3}_{1}x_{center}^{2}+  \\ 
\end{displaymath} 
\begin{displaymath} a^{4}_{1}x_{center}y_{center} + a^{5}_{1}y_{center}^{2} 
\end{displaymath} 
\begin{displaymath} \Delta x = x -x_{center} 
\end{displaymath} 

The terms $x_{center}$ and $y_{center}$ are the
central coordinates of the direct image. The coefficients ($a^{0}_{0}$,
$a^{1}_{0}$, etc.) are calculated in \citet{cal}.

\begin{figure}[t!] 
\epsscale{1.0} 
\vspace{0.2in} 
\plotone{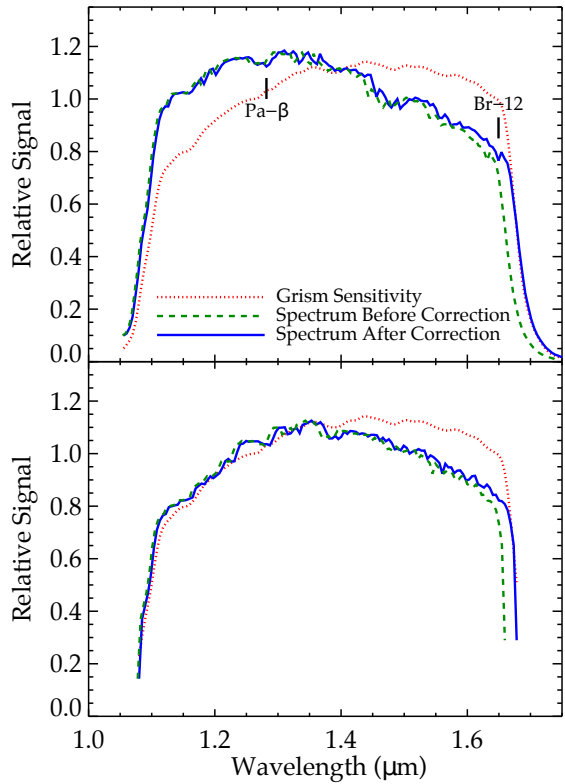}
\caption{Flat-field-corrected spectrum of TRES-2 (\emph{above}) and CoRoT-2
(\emph{below}). Each plot shows the WFC3 G141 grism sensitivity curve
(\emph{red, dotted line}) and the spectrum before (\emph{green, dashed line})
and after (\emph{blue, solid line}) the correction has been made to the
wavelength solution coefficients. The two hydrogen lines, Pa-$\beta$
(1.282\,$\mu$m) and Br-12 (1.646\,$\mu$m), the two lines in TRES-2 used to
adjust the wavelength coefficients, are also marked here. To get a normalized
spectrum, one must simply divide by the sensitivity curve.}
 \label{Fig1}
\end{figure}

In performing this calibration, we found that the calibrated grism
response (sensitivity) curve did not line up precisely in wavelength
space with the observed response (see Figure~\ref{Fig1}). We therefore adjusted the
coefficients empirically to obtain optimal agreement with the observed
grism response curve and with the wavelengths of two stellar hydrogen
lines (Pa-$\beta$ at 1.282\,$\mu$m and Br-12 at 1.646\,$\mu$m). These
adjustments yielded:

\begin{displaymath} a^{0}_{0} \rightarrow 0.997 \times a^{0}_{0}
\end{displaymath} 
\begin{displaymath} a^{1}_{0} \rightarrow 0.90 \times a^{1}_{0} 
\end{displaymath} 
\begin{displaymath} a^{0}_{1} \rightarrow 1.029 \times a^{0}_{1} 
\end{displaymath}

We used these adjusted values in the calibration presented in this
work and also successfully applied them to other data sets in this HST
program. Therefore, this empirical correction is not specific to this
target or these visits, and we in fact used another object in the program (TrES-2) to find the correction, as it was observed on the larger subarray and thus the observations include all 150 pixels of the spectrum.

\subsection{Flux Calibration}\label{sec:flux}

The flat field and sensitivity curve of the G141 grism on the WFC3 detector are
the two components of flux calibration, and both are wavelength-dependent.

For imaging observations, calwf3 applies the flat field to the data, but
flat-fielding of grism data must be done by the observer. STScI provides a
flat-field cube for the G141 grism. This cube is a four-extension FITS file, and
each extension is the size of the full WFC3 IR array. For a given pixel on the
data image with a given wavelength, the flat-field value for that pixel is given
by a polynomial function with coefficients defined by the values of the
flat-field cube extensions at the pixel's location.

This method is described in the aXe handbook \citep{axeman} and laid out with
the equations that follow. For a pixel at position (i,j), they define a
normalized wavelength coordinate, $x$: 
\begin{displaymath} x = \frac{\lambda -\lambda_{min}}{\lambda_{max} - \lambda_{min}} 
\end{displaymath} The parameters $\lambda_{min}$ and $\lambda_{max}$ are constants found in 
the flat-field cube header. The flat field value of a pixel (i,j) with normalized wavelength
coordinate $x$ is then a polynomial function in $x$: 
\begin{equation} f(i,j,x) = a_{0} + a_{1}x + a_{2}x^{2} + a_{3}x^{3} 
\end{equation} 
where $a_{0}$, $a_{1}$, $a_{2}$, and $a_{3}$ are the values at (i,j) 
in the zeroth, first, second, and third extension arrays in the flat-field cube file, 
respectively.  For both our $\alpha$ and $\beta$\ analyses, we apply the flat-field 
correction to the spectral box by dividing by the corresponding flat-field ``box," generated
pixel-by-pixel from the method above.

STScI also provides the wavelength-dependent sensitivity of the G141
grism. In Figure~\ref{Fig1}, we plot the scaled sensitivity curve
over a flat-fielded spectrum from a single exposure of TRES-2, along
with the two hydrogen lines used as reference points in adjusting the
wavelength calibration coefficients.

\subsection{Second, Overlapping Source}

\begin{figure} 
\epsscale{1.0} 
\vspace{0.2in} 
\plotone{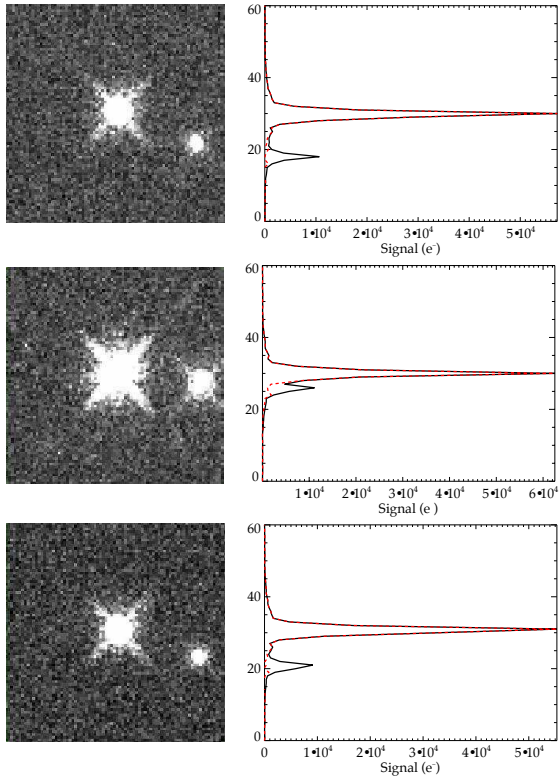}
\caption{\emph{Left:} The direct image of CoRoT-2 (brightest object, center of
each image) and the infrared source nearby in visits A (\emph{top}), B
(\emph{middle}), and C (\emph{bottom}). \emph{Right:} A vertical profile of the
first-order spectrum resulting from a horizontal dispersion of the light to the
right of the direct image for each of the three visits; the solid, black line is the original trace, while the dashed, red line is the trace after correction. The variation in degree
of overlap of the spectral trace is due to variation in the orientation angle of
the telescope between the visits, which changes the proximity of the second
source's spectrum to that of the target. The orientation angle was limited to the range
76$^{\circ}$\textless ORIENT\textless 166$^{\circ}$ by telescope operation parameters, and the actual angles were
93.9$^{\circ}$, 80.4$^{\circ}$, and 90.7$^{\circ}$, for visits A, B, and C.}
\label{Fig2} 
\end{figure}

CoRoT-2 has a companion star, so our analysis must remove or
correct for this second source.  The direct image of CoRoT-2 appears in
Figure~\ref{Fig2}, where the second, fainter source is evident. The proximity of
the second source in the image depends on the orientation angle of the
telescope, and varies between the three visits, but it is close enough to be of
concern for source contamination. The spectra overlap minimally in visits A and
C, but there is significant overlap in visit B, which has the lowest orientation
angle and thus the smallest distance between the two spectra of the three
observations. The orientation angles, which only vary a few degrees from each other, are reported in Table~\ref{tab:corot2obs}.

\subsubsection{Characterization} This second source is an infrared source, 2MASS
J19270636+0122577, but is just barely spatially resolved by the 2MASS observations. In the planet's discovery paper, \citet{corot} suggest it may be a late-K or M-type star, and \citet{companion}
identify it as a late K-type star. Both works posit that it may be gravitationally
bound to CoRoT-2.  We here address how to remove, or correct the effect, of this
second source from the flux of the CoRoT-2 system.  We have explored two
approaches.  Our $\alpha$ analysis removes the second source prior to extracting
the grism spectrum from the 2-D frames. Our $\beta$ analysis includes the second
source in the extracted grism spectra, and corrects the derived exoplanetary
spectrum after deriving that stellar spectrum.

The location of the second source allows us to generate its spectrum, albeit in
a limited wavelength range.  Its spatial offset results in losing the
long-wavelength end of its spectrum.  Comparing the partial spectrum to a grid
of Kurucz models shows general agreement with the findings of \citet{companion};
a temperature of 4000 K and surface gravity log(g) = 4.0, produces the best
agreement with the observed partial spectrum.  That corresponds to a late K- or
early M-type main sequence star.

\subsubsection{Removal}

In our $\alpha$ analysis, the strategy for removing the second source from
visits A and C is to determine the average spatial shape of the source's signal,
and scale and subtract it from each column of the spectral box. We fit a Gaussian
plus a second-order polynomial baseline to the spatial profile at each column of
the data. Averaging that fit over all exposures then approximates the signal
from the second source for a given column, after scaling the average to
represent the amplitude of the second source for each column.  The original
spectral trace -- the plot of wavelength-integrated flux versus spatial pixel --
appears in Figure~\ref{Fig2}, as well as the corrected spectral trace
(overplotted), showing significant improvement.

For visit B, the task is more difficult. The peaks of the two sources are
separated by just four pixels, compared to twelve and ten pixels for visits A
and C, respectively. The overlap leaves too few points to use a fitting
procedure to isolate and approximate the source. Instead, we use the descent of
the PSF on the opposite side of CoRoT-2 from the overlap of the second source,
and mirror the PSF column-by-column onto the side with overlap.  We subtract the
mirrored PSF, and fit a Gaussian column-by-column to the difference.  Averaging
that fit over all exposures, we approximate and remove the second source from
each column for visit B.

\section{Systematics: Characterization}\label{sec:systematics} 

Our data exhibit trends in the measured stellar intensity that are not
manifestations of physical stellar or planetary phenomena. Instead,
they represent tendencies of the detector to report signal counts that
are different from what actually fell on a given pixel.

We note that instrument-related systematic errors in WFC3 exoplanetary
spectroscopy are believed to be less severe than in NICMOS
observations \citep{gibsonnicmos, crouzet, deming13,
  swain13}. Nevertheless, instrument effects do exist in the WFC3
data, especially for observations taken before the advent of spatial
scan mode \citep{pm2} such as ours.  Some aspects of these
instrumental effects were discussed by
\citet{swain13}. \citet{berta12} reached nearly the photon limit in
their analysis of WFC3 G141 transit spectra of the super-Earth
GJ1214b, as did \citet{deming13} for two giant transiting exoplanets
with the implementation of the spatial scan mode. We will discuss the
analysis of the \citet{berta12} work and how we modified it for more
general purposes in \S\ref{sec:whitelight}.

\begin{figure*}[t!] 
\epsscale{1.0} 
\vspace{0.2in} 
\plotone{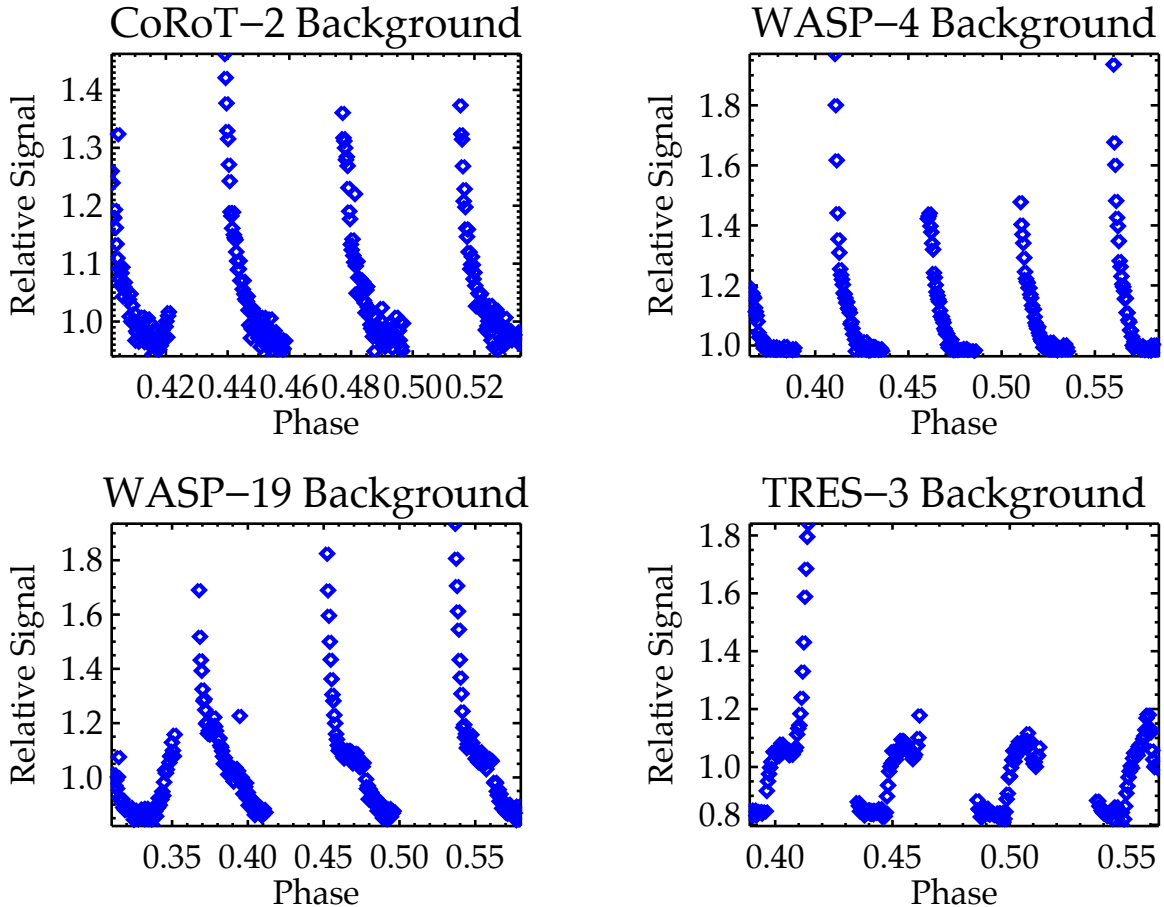}
\caption{The normalized signal measured from the background pixels over the
course of the observations. This systematic effect resets after each orbit of
the telescope (between orbits there is a gap in time as HST passes behind the
Earth). For most objects in our program, the effect is a smooth exponential
decrease, as shown here for CoRoT-2 and WASP-4 in the upper panels. For some
observations the shape is different, an irregular "S"-shape, as for WASP-19 and
TRES-3 in the lower panels. } 
\label{Fig3} 
\end{figure*}

We identify three primary manifestations of systematic error, and all
are patterns in intensity as a function of time. The first is a
continuous trend of the source's white light curve lasting the entire
length of a visit, during which the intensity gradually decreases with
time (or increases, in one case). This ``visit-long ramp" is linear in
nature (to within the errors), and continuous between orbits. Its slope varies widely between observations, not only among the CoRoT-2 visits, but among all in our HST program. Its strict linearity and variation even when separately observing the same star places it clearly in the category of instrumental effects rather than stellar modulations, but the exact cause is an open question. The
second systematic error feature is a decrease of intensity as a function of time, which repeats
for every orbit. This effect is apparent in the pixels {\it not}
illuminated by the source spectrum -- including the pixels we use for
the background subtraction -- and is shown for the examples of CoRoT-2
and others in Figure~\ref{Fig3}. For most objects in the program, the
effect shows a smooth, exponential decrease in the signal for these
pixels over the course of an orbit, though some observations show an
effect in more of an "S"-shape instead. Removal of the background, as
per the method described in \ref{sec:bg}, removes any discernible
presence of this effect, which allows us to conclude first that the
effect is isolated to the lowest-valued pixels, and second that we
need not perform further tasks to eliminate this orbit-long feature,
as the problem is solved by careful background definition and
subtraction. We find no definitive cause, though we suspect it may be
due to scattered light from the Earth's limb.

\begin{figure*}[t!] 
\epsscale{1.0} 
\vspace{.5in} 
\plotone{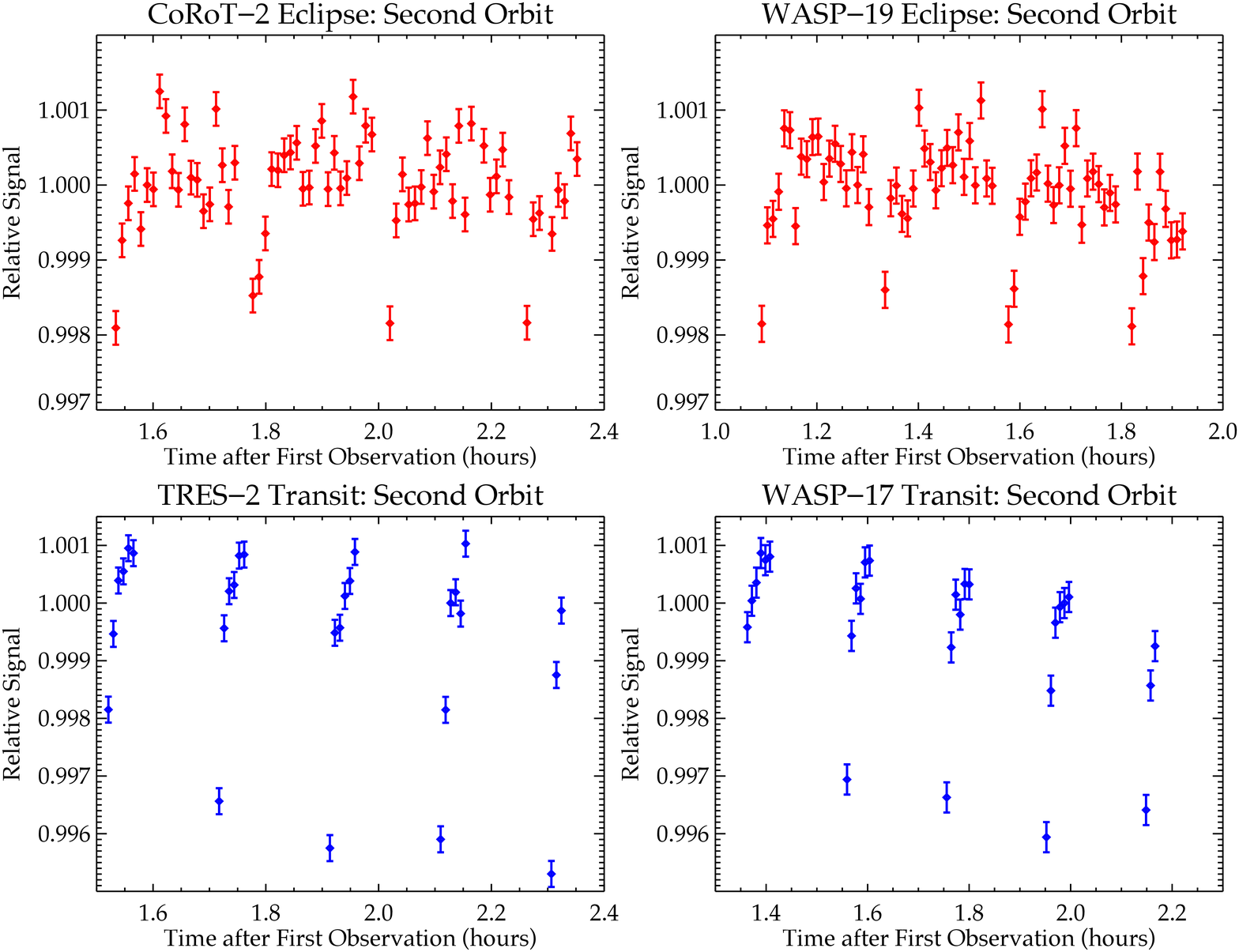}
\caption{Four examples of the systematic hook pattern. WASP-19 and CoRoT-2
(\emph{top}) were both observed on the smaller subarray, and have more exposures
in the pattern, a more subtle pattern, and less time in between iterations.
HAT-7 and TRES-2 (\emph{bottom}) were both observed on the larger subarray, and
have fewer exposures in the pattern, a more obvious pattern, and a much larger
gap in time between the observations.} 
\label{Fig4} 
\end{figure*}

The third example of systematic error is an increase in intensity of
the source's white light curve which occurs on a shorter time scale,
over the course of several exposures, and which repeats three or more
times in every orbit. We call this the "hook" \citep{deming13} because
of its characteristic shape, which is a steep jump for the first one
to three exposures and then a flattening\footnote{Some investigators
  call this effect a `ramp', but we advocate different terminology so
  as not to confuse it with the visit-long ramp, and also to
  distinguish it from the Spitzer ramp.}.  The hook appears to a
varying degree in all of the observations, and produces a significant
distortion in the data. Examples of the hook within a single orbit of
observations for four different objects are shown in
Figure~\ref{Fig4}. The reset of this pattern corresponds with the time
when the data stored in the WFC3 buffer are sent to the solid-state
recorder on the spacecraft. This causes a short break in observations,
and also resets the detector.  Neither of the other two primary
systematic effects appear to have any dependence on times of data
transfer.

The hook pattern is of similar shape in all sets of observations in
the program, but the parameters of its manifestation, e.g., length of
time, number of exposures, number of iterations, vary from object to
object. Figure~\ref{Fig4} shows examples of the pattern in four
different objects; the shape of the pattern is similar, but the
amplitude of the hook and the time between buffer dumps (and thus the
number of exposures and total time of each hook)
varies. \citet{swain13} concluded that it was most significant for the
$512\times512$ subarray.  We concur that it is often prominent at
$512\times512$, and it is considerably steeper for longer-duration
patterns on $512\times512$, but we detect it in other subarrays also.
The prominence of the pattern correlates with brightness of the star.

While the visit-long slope appears to be linear, both the orbit-long and hook
effects are exponential in shape, and therefore each begins as a very strong
effect and then becomes nearly indiscernible in the final exposures of each
hook pattern.

There are further apparent systematic effects seen in the first orbit of every
observation; they are most likely due to telescope settling and readjusting to a
new pointing, and do not have a consistent pattern.  Therefore we discard the
first orbit once we begin applying corrections to the systematic effects for the
purpose of calculating the wavelength-integrated transit depth, and the spectrum
of the planet.  Since the eclipse of CoRoT-2b is covered by three visits, loss of the first
orbit is only a minor perturbation for our analysis.

\subsection{Persistence Correction}\label{sec:persistence}

One potential cause of the hook effect is detector persistence, the
phenomenon in which trapped charge in an exposure is slowly released
in following exposure(s) to produce a falsely increased signal
detection \citep{smith08}. STScI publishes persistence models and even
predictions of persistence for a given exposure based on the exposures
prior to it. The predictions are for an additive effect, and the data
product for a given exposure is an image array the size of the
original exposure, but with each pixel value equal to the the
predicted persistence, so the correction is simply to subtract the
corresponding pixel values. The persistence is low for the first
exposure, but jumps up quickly and remains at a higher value until the
time of the data transfer, when it, too, resets.  The additive
correction as given by STScI do decrease the severity of the hook, but
they do {\it not} entirely remove the hook, and we conclude that the
hook is a combination of a additive and {\it multiplicative}
effect. This will justify our methods of correction outlined and
examined in the sections that follow. We have made the STScI
persistence correction in our $\alpha$ analysis.  Our $\beta$ analysis
ignores additive persistence, as do most WFC3 exoplanetary
investigations published to date.

\subsection{Pixel-by-Pixel Evaluation of the Hook}\label{sec:hookcorrection}

\begin{figure}[t!]
\epsscale{1.0} 
\plotone{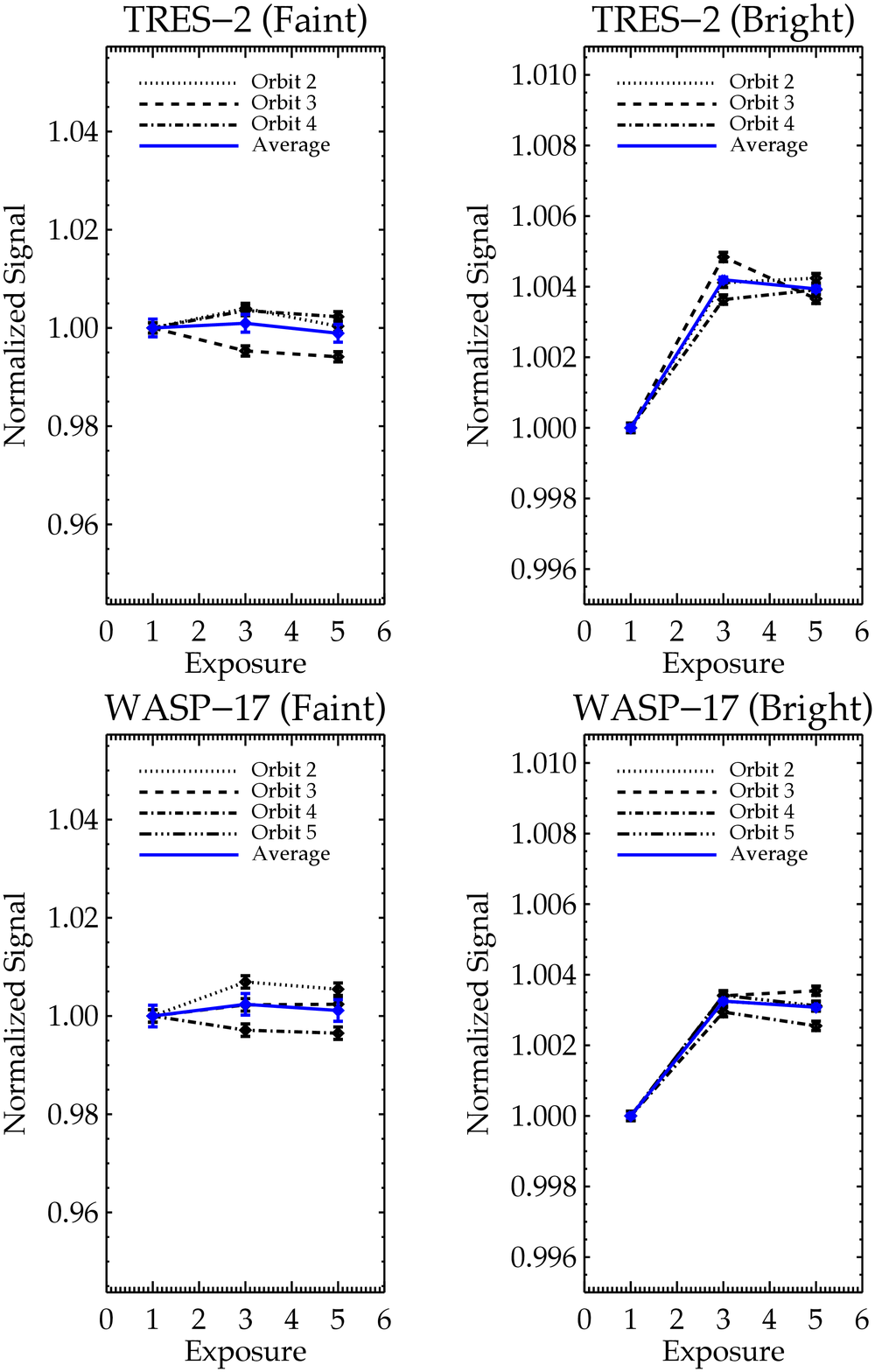}
\caption{Examining the average shape of the hook pattern for two sets of
observations.  We calculated the total flux in the spectral box at the
beginning, middle, and end of the pattern, averaged over all the iterations
within an orbit, and then plotted the average normalized to the first average value. For
each object shown, the pixels in the spectral box have been split in half about the median value: the
faint half and the bright half, and then plotted separately. As is apparent with
this split, the fainter pixels are not affected by whatever causes the pattern,
while the brighter half are.} 
\label{Fig5} 
\end{figure}

\citet{berta12} demonstrated that the hook is more prominent at
high exposure levels.  We have investigated the amplitude of the hook as a
function of the per-pixel exposure level, and other parameters, and we seek
quantitative relationships.  For each pixel, we average the change in signal
level over the multiple iterations of the pattern within one orbit, and then
examine the change as a function of time within the pattern, flux of the pixel,
and location of the pixel on the detector.

The average shape of the hook for two objects in the program can be seen in
Figure~\ref{Fig5}. The normalized signal is shown against the exposure number
within the pattern. For each visit, the pixels have been split between those
with flux below the mean and those with flux above the mean. This is done to
confirm that the existence of the hook does indeed depend upon the flux of the
pixel.

\begin{figure*}
\epsscale{1.0}
\plotone{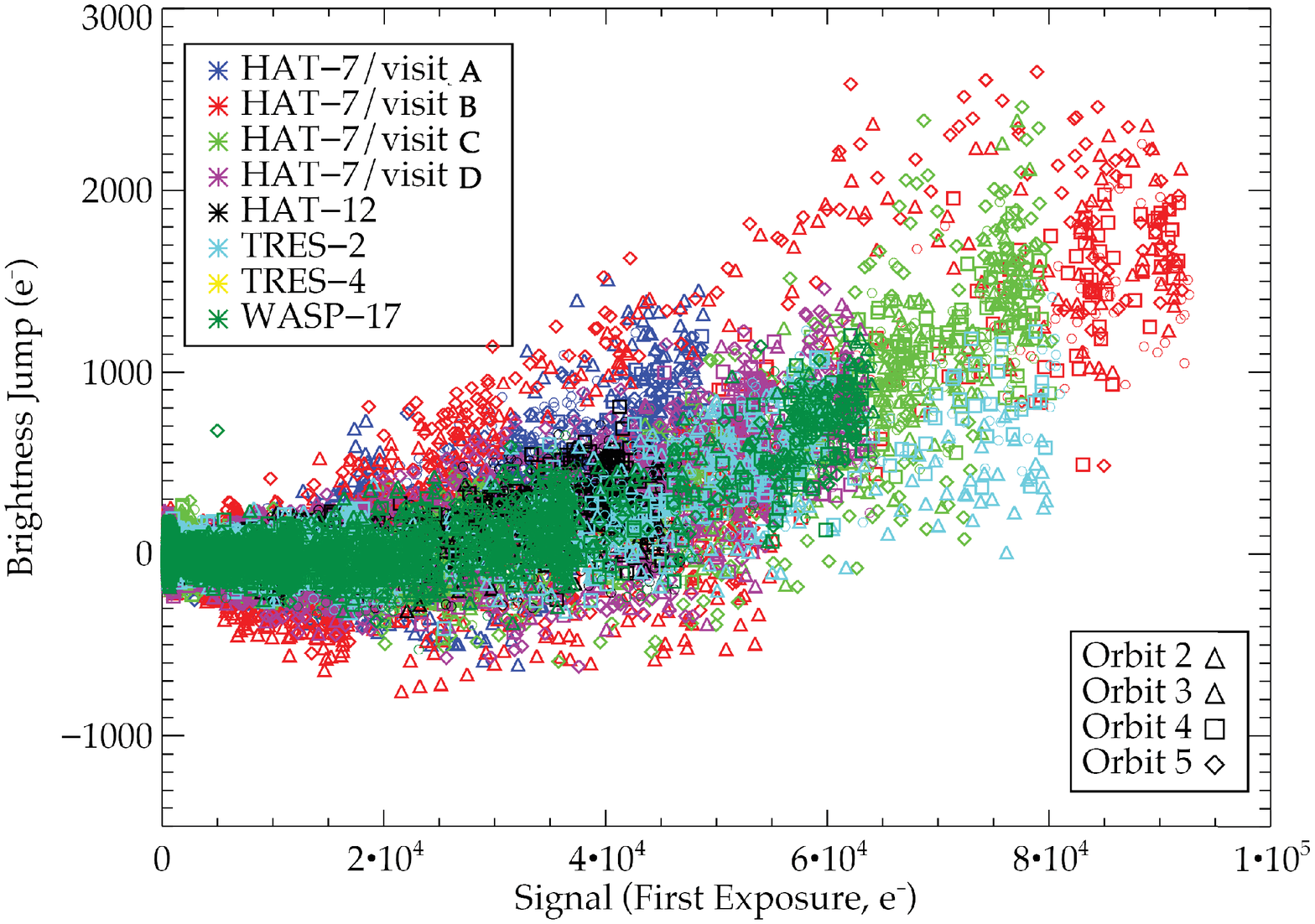}
\caption{A quantification of an additive effect from the detector for a selection of objects. The hook pattern
repeats multiple times in each orbit, and each visit has 4-5 orbits. For every
orbit in every visit, coded by symbol and color, we have averaged the increase
in measured flux from the first to the second-to-last exposure in each pattern
for each pixel in the spectral box. This is plotted against the initial flux of
each pixel in the first exposure of a hook. The increase is clearly dependent
upon the flux level, and does not become apparent (on average) until a signal of
about 30,000 electrons. The legend shows which visit corresponds to which color.
The relation between initial flux and flux jump appears to be steeper for longer
pattern times.}
 \label{Fig6} 
 \end{figure*}

Figure~\ref{Fig6} shows this dependence of the additive change on the flux of
the pixels, where every pixel has been plotted by its initial flux and "jump" in
electrons between the first exposure and the last exposure in the pattern. The
jump is statistically insignificant below a certain original pixel value, but
shows a reliable parabolic rise starting around 30,000 electrons. The scatter is
nevertheless remarkably large, which ultimately means that we cannot depend on a
unique quantitative relation to correct this effect.

In principle, the hook could be removed by using Figure~\ref{Fig6} to predict
the magnitude of the jump for a pixel given its initial flux in the first
exposure of the pattern, and thereby correct each pixel in each image.  We
attempted such a correction, and it does remove the obvious appearance of the
hook pattern, but it leaves the data with much more scatter than is acceptable,
due to the wide variations seen in Figure~\ref{Fig6}.

\begin{figure}[t!]
\includegraphics[scale=0.35,angle=90]{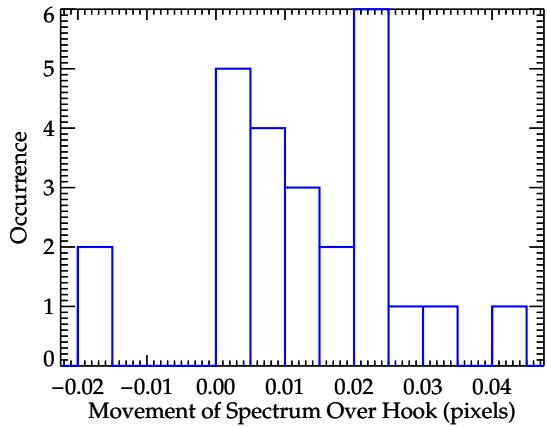}
\caption{To measure any possible dependence of the hook effect on pixel row, we
compared a Gaussian fit to the spectrum between the first and last observations
of the hook for all iterations in the objects observed on the 128$\times$128
subarray. We show here that the location of the maximum point of the spectrum
(the peak of the Gaussian) typically moves, and typically moves in the same
direction, over the course of the hook. This indicates that the hook pattern has
a row dependence.} 
\label{Fig7} 
\end{figure}

We also examined the amplitude of the hook as a function of position
on the detector. We find no correlation in column (wavelength) space,
but some correlation with the slope of the hook pattern and the row on
the detector, i.e., how far a given row is from the spatial center of
the spectrum.  This correlation does seem to strongly depend on which
subarray we used. Especially in the case of the 128$\times$128
subarray, the slope of the trend is more positive for the rows of
pixels below the central peak of the spectrum (in the direction
perpendicular to dispersion), while the slope is less positive for
those rows above the central peak.  This correlation is weaker for the
512$\times$512 subarray, but still discernible.  Figure~\ref{Fig7}
shows the correlation for the smaller subarray by demonstrating the
shift in the spatial center of the spectrum between the starting and
ending frames of the hook. Our finding that the nature of the hook
depends on the row of the spectrum may be a significant clue to the
nature of this effect.  Reading the detector involves addressing the
pixels by row, and it is conceivable that the hook is related to the
manner in which the detector is addressed and sampled. We conclude
that the effect in Figure~\ref{Fig7} cannot be explained by anything
like telescope drift. The trend featured in Figure~\ref{Fig7} is
correlated with the hook and therefore the transfer of the detector
buffer, a task performed with no relation to telescope motion.

\section{White-Light Eclipse Curve}\label{sec:whitelight}

We wish to produce a time series of the wavelength-integrated (`white light')
signal measured from CoRoT-2 in order to determine the amplitude and central
phase of the secondary eclipse.  This will yield the total signal from the
planet over the G141 bandpass, while the spectrum that we calculate in
Sec.~\ref{sec:spectrum} will distribute that signal as a function of wavelength.
We begin with the light curves for the three visits shown in Figure~\ref{Fig8},
then we correct these light curves to remove the instrumental systematic
effects, and we combine the three visits to form a single eclipse curve as a
function of orbital phase.

\citet{berta12} successfully removed systematic effects from their dataset. The
steps of their {\bf divide-oot} method for correcting a transit/eclipse curve are as
follows, assuming a five-orbit set of observations, with orbits three and four
in transit: \begin{verse} 
  1. Ensure that all orbits have the same number of
exposures. The fifth orbit usually has fewer exposures than orbits two, three,
and four, so simply repeat the last element to make up the difference. Since the
hook pattern is flatter at its end, this is a reasonable approximation. \\ 
  2.Create an average out-of-transit orbit by simply averaging orbits two and five.\\ 
  3. Divide each orbit (two, three, four, and five) by the average orbit. \\ 
  4.Remove the artificial elements that were added in the first step. \\ 
  5. Fit a line to the second and fifth orbits, as there is still usually a hint of the
visit-long ramp. Divide by the linear fit to normalize the data in units of the
stellar flux.\\ \end{verse} 

This method should yield an acceptable eclipse curve
with out-of-transit flux normalized to unity. Application of {\bf divide-oot} to
objects in our HST program 12181 proved successful only in some cases
\citep{ranjan13}. A modification of the method will be explained below.

\subsection{Modified {\bf divide-oot}}\label{sec:mod_oot} 
We observe CoRoT-2 in four orbits per visit, but each visit contains
at most one orbit that is completely in-eclipse (when the planet does
not contribute), and each visit contains the virtually unusable first
orbit. For this reason and due to our significantly lower
signal-to-noise ratio than for the \citet{berta12} planet's observations -- the GJ1214b transit depth is two orders of magnitude larger than the depth of the CoRoT-2b secondary eclipse in the same waveband and on the same grism -- our CoRoT-2 data are not well-suited for the {\bf divide-oot} method \emph{per se}.  Another issue with CoRoT-2 is the severity of the visit-long ramp, which causes trouble when trying to average
pattern shapes before removing the ramp.  Therefore, instead of
dividing by an average \emph{orbit}, we elect to divide by an average
\emph{pattern}, defined both by the occurrence of a buffer dump and
through visual assessment, and we proceed as follows: \begin{verse}
  1. Identify the patterns that are out-of-eclipse. Divide the entire
  white-light curve by the median of the out-of-eclipse exposures from
  a single, early orbit (usually orbit 2). This normalizes the curve
  to unity. \\ 
2. Fit a line to the out-of-eclipse patterns, but
  exclude all points below intensity level 0.997. These outliers are
  due to the hook effect, and would bias the visit-long slope
  correction. \\ 
3.  Divide by the fitted curve to re-normalize to
  unity. \\ 
4. Create an average pattern by averaging the
  out-of-eclipse patterns. \\ 
5. Divide each occurrence of the pattern
  in the entire white-light curve by the average pattern. \\ \end{verse} 

\begin{figure*}[t!]
\includegraphics[scale=0.7,,angle=90]{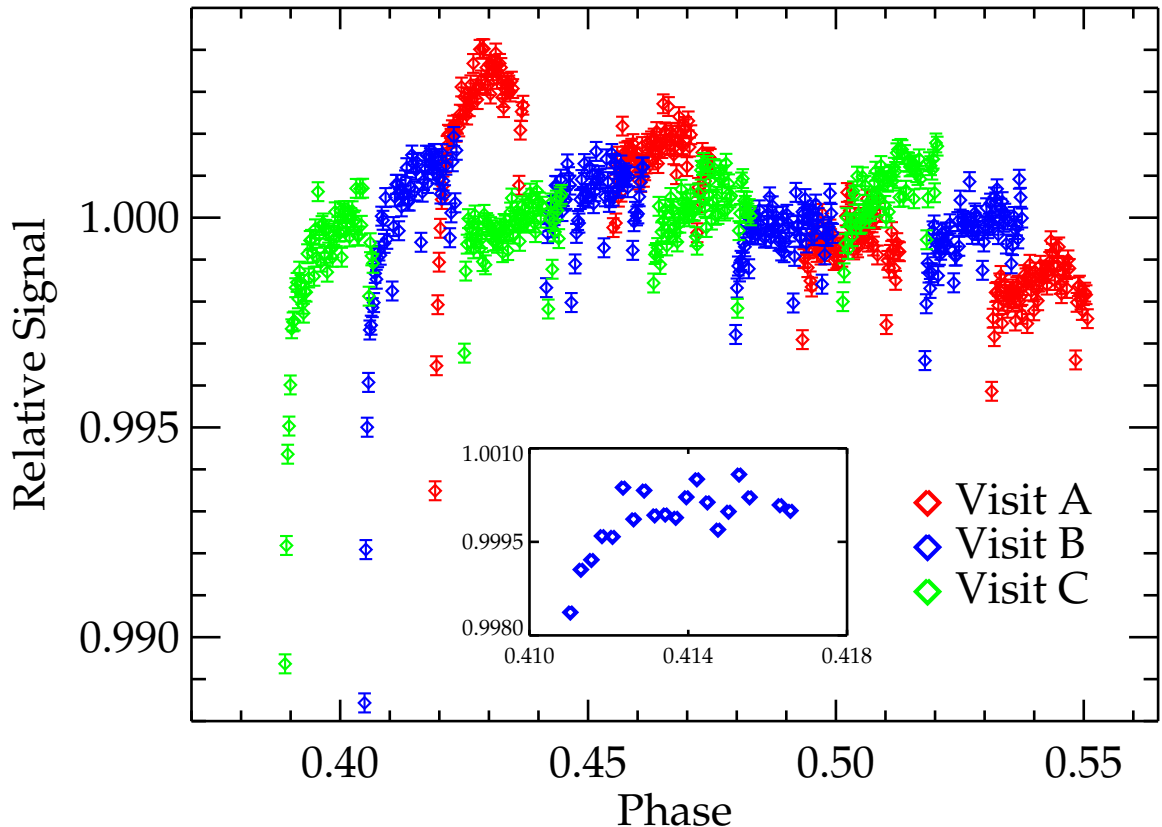}
\caption{Phase plot of the wavelength-integrated flux from the three visits of CoRoT-2 before any
corrections to the systematics have been applied. Visits A, B, and C are shown
in red, blue, and green, respectively. \emph{Inset:} An example of an "average pattern," of the characteristic hook shape, corresponding to visit 23 and calculated by the modified {\bf divide-oot} method described in \S\ref{sec:mod_oot}. This pattern is calculated after removal of the linear visit-long ramp.}
\label{Fig8} 
\end{figure*}

This creates a vast improvement in the data, with a significant reduction in the presence
of systematic effects. We are also able to utilize the later patterns of the first orbit, rather than discarding it completely, as the problems presented by settling or other effects of unknown origin diminish significantly after one to two iterations of the pattern. An average pattern is plotted in the inset of Figure~\ref{Fig8}, and the corrected data are shown in comparison to
our best-fit eclipse curve in Figure~\ref{Fig9}.

\subsection{White-Light Eclipse Amplitude}\label{sec:eclipsedepth} 
With the corrected data in hand after applying our modified {\bf divide-oot}
procedure, we fit an eclipse curve using the data from our $\alpha$
analysis.  We calculate the shape of the theoretical eclipse curve
using expressions from \citet{transitorb}, with orbital parameters
from \citet{alonso09}, except for the orbital period where we adopt
the updated value from \citet{sada}. In fitting the data, we vary only
the central phase and amplitude of the eclipse, the latter by scaling
the amplitude of the theoretical curve.  We perform the fit using two
$\chi^2$-minimization methods.  First, we implement a
Levenberg-Marquardt algorithm to vary the eclipse amplitude and
central phase simultaneously, to find the global minimum in $\chi^2$.
Second, we vary the central phase incrementally from 0.49 to 0.51 in
steps of $10^{-5}$.  At each step, we calculate the best-fit eclipse
amplitude at that phase in closed form, using linear least-squares.
Cycling through the range of trial central phases, we again find the
global minimum $\chi^2$.  Results from the two methods were in
excellent agreement.

\begin{figure}[t!] 
\epsscale{1.0} 
\plotone{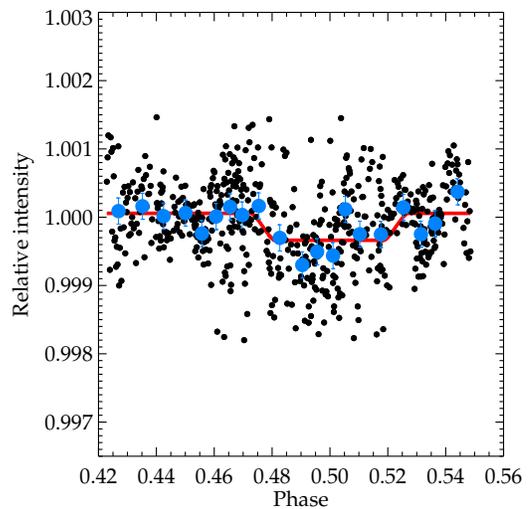}
\caption{Wavelength-integrated light curve of CoRoT-2 after correction of the
hook and visit-long ramps as described in section \S\ref{sec:mod_oot}. The
best-fit secondary eclipse curve is overplotted in red. The large points in blue
represent averages over bins of 0.0063 in phase, about 15 minutes in time.  The
fit was performed on the actual data (black points); the binned data are shown
merely for reference.} 
\label{Fig9} 
\end{figure}

We find a best-fit eclipse depth of $395^{+69}_{-45}$\,ppm (parts per million);
the fit is shown in Figure~\ref{Fig9}.  The reduced $\chi^{2}_{red} = 6.60$; as it was calculated estimating the error to be Poissonian, the ideal scenario, this $\chi^{2}_{red}$ value indicates that the achieved per-point scatter is 2.6 times the photon noise. The error level, and the appearance of Figure~\ref{Fig9}, suggests that red noise remains in the data, in spite of our modified {\bf divide-oot} procedure.  To
verify the presence of red noise, we binned the residuals from the best-fit
eclipse over $N$ points per bin, and calculated the standard deviation of the
binned points, $\sigma_{N}$.  We solve for the slope of the relation between
$\log(N)$ and $\log(\sigma_{N})$ using linear least-squares.  Poisson noise will
produce a slope of $-0.5$, whereas we find a slope of $-0.33\pm0.03$ for the
Figure~\ref{Fig9} data, confirming the presence of red noise.

\begin{figure}[t!]
\epsscale{1.0} 
\hspace{-0.4in}
\plotone{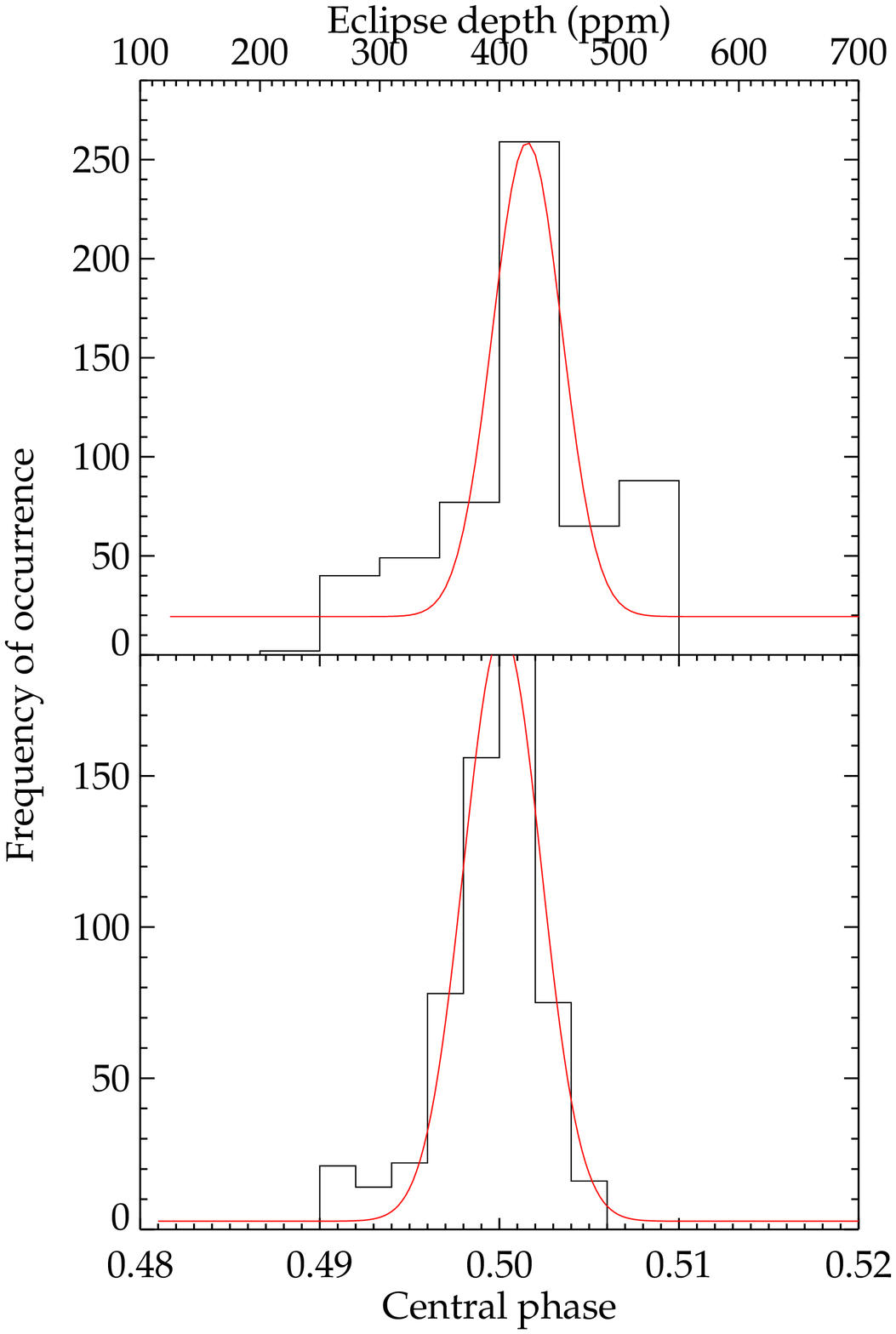}
\caption{Error analysis for the amplitude and central phase of the white light
eclipse. The frequency of occurrence is based on a total of 580 residual
permutations.  Upper panel: histogram of eclipse amplitudes in parts-per-million
for the residual permutation error analysis of the eclipse amplitude.  Lower
panel: histogram from the residual permutation error analysis of the central
phase of the eclipse.} 
\label{Fig10} 
\end{figure}

Given the presence of red noise in the white light eclipse data, we assign
errors to the best-fit eclipse parameters (eclipse amplitude and central phase)
using the residual permutation (``prayer-bead") method \citep{bouchy05,gillon07}. 
Figure~\ref{Fig10} shows histograms of the results for the best-fit amplitude
and central phase, based on the residual permutation fits.  For reference, we
fit Gaussians to these histograms.  A Gaussian is a reasonable approximation to
the central phase histogram, but the eclipse amplitude histogram has a higher
central peak, and lower wings, than does a Gaussian.  Our adopted errors are
equivalent to the $\pm1\sigma$ points in the histograms, in the sense that
$15.8\%$ of the histogram area lies beyond each quoted $1\sigma$ value ($31.6\%$
considering both ends of the range).

\subsection{Eclipse Central Phase}\label{sec:eccentricity}

Our best-fit eclipse is centered at a phase of $0.4998\pm0.0030$.  The
light-travel time across the orbit is 28 seconds. The central phase
for a circular orbit would be 0.50019, consistent with our result,
within our errors.  \citet{gillon10} found the eclipse to occur
slightly earlier than expected for a circular orbit, at phase
$0.4981\pm0.0004$. \citep{drakespitzer} found a central phase of
$0.4994\pm0.0007$, weakly supporting the result from \citet{gillon10}.
The low signal-to-noise -- due to the shallower secondary
eclipse at shorter wavelengths  -- of the eclipse in the WFC3
bandpass contributes to a relatively large error level for the central
phase (approximately 4 to 8 times larger than the Spitzer errors).
Although we find good agreement with a circular orbit, we cannot
exclude the result of \citet{gillon10} who concluded that the orbit is
slightly eccentric.

\section{Calculation of the Eclipse Spectrum}\label{sec:spectrum}
\citet{berta12} used his {\bf divide-oot} method for GJ\,1214b to derive the
depth of transit as a function of wavelength, i.e., the transmission
spectrum.  In principle that method is applicable to exoplanetary
spectra at secondary eclipse, but we use an alternate technique.  We
have at most one in-eclipse reference orbit (when the planet does not
contribute) per visit.  Moreover, CoRoT-2 is a relatively faint star
(V=12.6, H=10.4).  In the faint-source limit, dividing
single-wavelength data by a single reference orbit would increase the
random noise in the quotient to an unacceptable degree, because we are
photon-starved.  To obtain the spectrum of the planet, we utilize the
differential method described by \citet{deming13} and explained below.  We apply this
method to data from both our $\alpha$ and $\beta$ data analyses,
finding consistent results.  

A by-product of this method is a time-dependent scaling factor
obtained by fitting a template spectrum (see below).  This scaling
factor is an excellent proxy for the white light eclipse, and we find
consistent results between the modified {\bf divide-oot} and differential
methods when calculating that white light eclipse.  That comparison
also served to verify that our $\alpha$ and $\beta$ analyses produce
consistent values for the white light eclipse depth.

\subsection{Beyond {\bf divide-oot}: the Differential Method}

The differential method is intended to exploit the characteristics of
the systematic hook pattern in order to cancel it, while also
correcting for the effects of jitter in wavelength over time.  The amplitude of
the hook is a function of the flux level in the affected pixels
(\S\ref{sec:systematics}). The procedure of the differential method, in its simplest form, is to therefore extract the intensity in each column of each grism image, and divide that intensity by the
wavelength-integrated intensity in the entire spectrum observed at
that time.  In other words, ratio the intensity in a given column
on the detector (after subtracting the background, and integrating
over rows) to the sum of all columns, and we repeat this process for the grism image at each orbital phase $\phi$.  This ratio adds minimal noise, because the precision of the wavelength-integrated spectrum is much better than the precision of a single wavelength.  Moreover, the ratio
should be effective in removing the hook, as long as the wavelength
used in the numerator is not too close to the edges of the grism
response, where the intensity rolls-off to much smaller values, as
does the hook (\S\ref{sec:systematics}).  The observed grism spectral
intensity varies only modestly (Figure~\ref{Fig1}) over the
1.1-1.7\,$\mu$m range of our analysis. Thus, dividing a single
wavelength by the sum of all wavelengths is a comparison of similar
intensity levels, so we expect much of the hook pattern to cancel, and
this expectation is met by the actual data (see below).

The differential method also removes the white-light eclipse. Specifically, the
eclipse shown on Figure~\ref{Fig9}, by summing over wavelengths, will identically
cancel.  However, wavelength-to-wavelength {\it variations} in the eclipse depth
will be preserved.  We call these differences {\it differential depths} and we
derive them either positive or negative, by fitting to the wavelength-ratioed data. 
We then add the depth of the white-light eclipse, reconstructing the full
emergent spectrum of the planet at eclipse.  

In actual practice, the implementation of this differential method is more complex than the simple
division implied above.  We do not explicitly divide by a wavelength integral;
we use an equivalent but more subtle procedure that we now describe.

\begin{figure} [t!]
\epsscale{1.3} 
\plotone{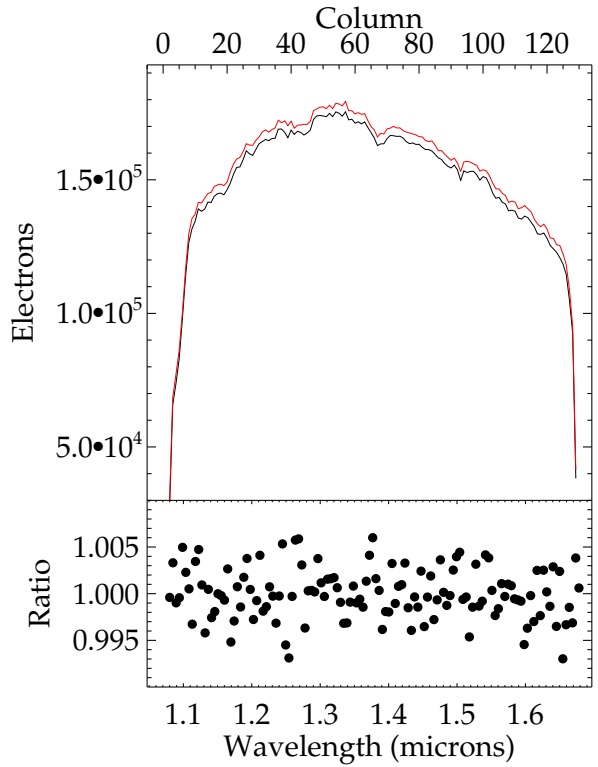}
\caption{Top panel: Spectrum of the star + planet (black line) at a randomly
selected time during visit A, compared with a best-fitting 'star only' spectrum
(red line) constructed as an average of all of the in-eclipse spectra during
visit A.  (These spectra are prior to the smoothing that we employ.) The
star-only spectrum was shifted in wavelength and scaled in intensity to provide
the best fit to the star+planet spectrum (see text, however for this figure an
additional 2\% shift in intensity was added so that the two lines do not
overlap).  Bottom panel: ratio of the star+planet spectrum to the shifted and
scaled star-only spectrum.  The scatter (0.00245) is dominated by the photon
noise of the spectrum in the numerator of the ratio.} 
\label{Fig11}
\end{figure}

We must account for possible wavelength shifts in each grism spectrum.
Wavelength shifts have two effects.  First, a shift of the spectrum changes the
intensity in a given column because the grism response varies with wavelength. 
Second, a shift in the spectrum changes the range of wavelengths sampled by a
given column of the detector.  We find that the wavelength shifts are of order
0.02-pixels, and they vary within an orbit, but tend to reset and exhibit a
similar pattern in subsequent orbits.  Given this magnitude of shifts, the
second effect mentioned above - a perturbation to the wavelength assigned to a
given column - has negligible effect.  We therefore ignore the wavelength
perturbations per se, and we use the wavelength scale from the calibration
described in \S\ref{sec:wave}. However, the first effect (changes in grism
response with wavelength) is important, and we deal with it as follows: \begin{verse}
  1. For each visit, form a ``template" spectrum of the star alone by summing the in-eclipse (planet hidden) spectra. Denote this spectrum  by $S_{x}$, where $x$ is the column coordinate on the detector.\\
  2. Fit the template to each individual spectrum by re-sampling, shifting
(in steps of $10^{-4}$ pixels), and scaling $S_{x}$ in intensity using linear
least-squares. Perform this least-squares fit over a large range of shift values ($\pm 0.1$-pixels) and choose the shift that exhibits the best fit as judged by the standard deviation
of the ratio. \\ 
  3. Each individual spectrum, $P_{x}$ at orbital phase $\phi$, matches a version of $S_{x}$ with a scaling factor $a$: $aS^{'}_{x}+b$. The prime marks the change in intensity due to the shift in $x$, and the
zero-point constant $b$ is negligibly small. \\
  4. Form the ratio $R^{\phi}_{x} = \frac{ P_{x}}{ aS^{'}_{x}+b }$. \\
\end{verse} 
An example of this basic process of shifting and fitting the template spectrum, for a randomly selected spectrum in visit C, is illustrated in Figure~\ref{Fig11}. However, our actual analysis adds an additional step in
order to deal with the undersampling of the stellar spectrum as discussed by
\citet{deming13}. Between steps 3 and 4 above:\begin{verse}
  3.5. Smooth all of the spectra using a Gaussian kernel with FWHM = 4 pixels.\\
  \end{verse}   
The choice of pixels (columns in wavelength) is dictated by the tradeoff between
suppressing the undersampling, and preserving the spectral resolution.

The wavelength integrals of $P_{x}$ and $aS^{'}_{x}+b$ are closely equal because
of the fitting process that matches them.  Moreover, the shape of $S_{x}$ is
constant over a visit, i.e., its value at any single wavelength, relative to
its wavelength integral, is constant. Hence the point-by-point division
described above is conceptually equivalent to dividing a single wavelength
(equivalently, $x$-value) in $P_{x}$ by the wavelength integral of $P_{x}$. 
However, our procedure has the advantage that we do not have to re-sample any
spectra wherein the potential signal is present, or where the reference stellar
spectrum is changing.  Hence we introduce no extra noise in this process, while
also correcting for wavelength jitter in the spectrum.

\subsubsection{The Spectrum of CoRoT-2b Using the Differential Method}
\begin{figure} [t!]
\epsscale{1.0}  
\plotone{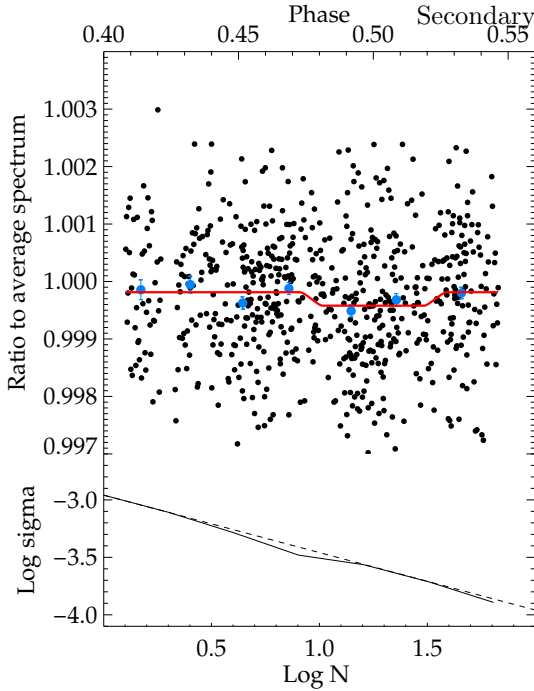}
\caption{Top panel: Differential eclipse at a single randomly-selected
wavelength ($\lambda = 1.551\,\mu$m).  Bottom panel: Log of the observed
dispersion (solid line) for bins of N points, versus log N. The dashed line
shows the relation expected for an inverse square-root dependence, as per photon
noise.} 
\label{Fig12} 
\end{figure}

Performing the procedure described above yields a set of ratio values
$R^{\phi}_{x}$ for each visit.  We now combine visits as follows: \begin{verse}
  1. For each column of the detector $x$, fit a straight line to the $R^{\phi}_{x}$, where
the independent variable in the linear fit is phase $\phi$, and then divide by that line. \\ \end{verse}
Dividing each visit by the linear fit removes any slight slopes that are present in each visit (as described by \citealp{berta12} and \S\ref{sec:systematics}) and places all three visits on a common scale. \begin{verse}
  2. Fit an eclipse curve to the combined $R^{\phi}_{x}$ at each $x$, holding the central phase fixed at $0.5$ for the eclipse fit, solving only for the depth. \\
  3. Use the wavelength calibration to associate a wavelength with each column $x$; $R^{\phi}_{x}$ becomes
$R^{\phi}_{\lambda}$. \\ \end{verse} 
The wavelength scale is sufficiently similar for each of the visits that we associate visit-averaged wavelengths with each $x$. The upper panel of Figure~\ref{Fig12} shows the result of fitting an
eclipse curve to the visit-combined $R^{\phi}_{\lambda}$ at a randomly-selected
wavelength.  Because the white-light eclipse has been removed by the process
used to generate the $R^{\phi}_{\lambda}$, the differential eclipse depth at
individual wavelengths can be either positive or negative depending on whether
the intensity of the exoplanetary spectrum is greater or less at that wavelength
compared to the average over the band defined by the grism response. Note that
the scatter in the individual points on Figure~\ref{Fig12} is large compared to
these differential eclipse depths.  However, the precision of the differential
eclipse depths is much better than the single-point scatter in $R^{\phi}_{\lambda}$,
and we also average adjacent wavelengths to derive spectral structure in the
exoplanetary spectrum (see below).

As the final step, \begin{verse}
  4. We add the white-light eclipse depth (0.000495, \S\ref{sec:eclipsedepth}) to the differential eclipse depths, and thereby derive the planet-to-star contrast versus wavelength. \\ \end{verse}
This emergent spectrum of the planet is illustrated on Figure~\ref{Fig13},
from both our $\alpha$- and $\beta$-analyses.  The upper panel shows
the values for individual wavelengths, i.e., single columns of the
detector, and the lower panel bins the results in bins of width
0.05\,$\mu$m (4 columns).

\subsubsection{Errors}\label{sec:errors} We have estimated the errors on the
differential eclipse depths using two methods.  For both methods, we
remove the fitted differential eclipse and examine the properties of
the point-to-point scatter (Figure~\ref{Fig12}, top) for each
wavelength.  First, we bin these points using bin widths of 2, 4, 8,
16, 32 and 64 points, and we calculate the scatter in those binned
values.  For Poisson noise, we expect that the scatter as a function
of bin size $\sigma (N)$ will decrease as $N^{-0.5}$.  An example of
the measured relation at a single randomly-chosen wavelength is shown
in the lower panel of Figure~\ref{Fig12}, where the dashed line is an
extrapolation from the single-point scatter using an exponent of
$-0.5$, and the solid line is what we calculate from the actual data.
These differential data are nearly photon-limited at almost all
wavelengths, and $\sigma (N)$ decreases very close to $N^{-0.5}$.  We
write $\sigma (N) = a \sigma (1) N^{b}$ and we solve for $a$ and $b$.
We then use that relation to calculate the expected precision for the
aggregate in-eclipse points and the aggregate out-of-eclipse points,
and we propagate those errors to calculate the error on the
differential eclipse depths.

As a check on the above error calculation, we also derive the precision of the
differential eclipse directly using the residual-permutation method
\citep{bouchy05}. Removing the best-fitting differential eclipse, we permute the
residuals sequentially and add them back to the best-fit eclipse curve to make
new data.  Fitting to these re-cast data for all possible permutations (580 of
them), we calculate the dispersion in the resultant differential eclipse depths.
 On average, we find that this produces excellent agreement with the first
method described above.  For our final spectrum and errors, we bin the results -
and propagate the errors - to the same resolution (4 columns, 0.05\,$\mu$m) that we used as a
smoothing kernel in the wavelength jitter correction. 

\begin{figure}[t!] 
\epsscale{1.0}  
\plotone{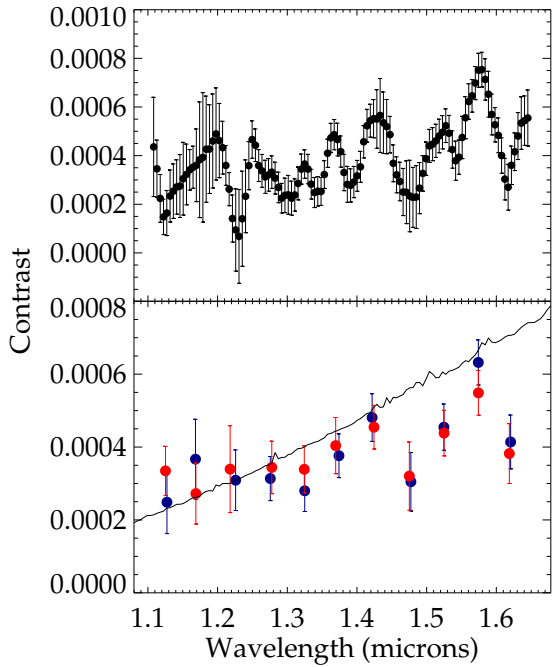}
\caption{Top panel: Eclipse depth (as planet/star contrast) versus wavelength
for the eclipse of CoRoT-2. Results from each detector column are plotted (from
our $\alpha$ analysis), so the smoothing used in the wavelength shift process
creates the appearance of autocorrelation. Bottom panel: Spectra of CoRoT-2b
from our $\alpha$ (red points) and $\beta$ analysis (blue points), binned to
0.05\,$\mu$m (4 column) resolution.  The line is a 1788K blackbody for the
planet.} 
\label{Fig13} 
\end{figure}

Figure~\ref{Fig13} shows the exoplanetary spectrum from our analyses
at single-column resolution (top panel, only $\alpha$ results for illustrative purposes), and binned to a wavelength resolution of 0.05\,$\mu$m (bottom panel).  The error bars on the
$\alpha$ binned spectrum in Figure~\ref{Fig13} are 77~ppm on average,
which is 25\% greater than the photon noise.  From our
$\beta$-analysis, the binned spectrum is similar, and the errors
average to 73 ppm (18\% greater than the photon noise).  The values of
our binned spectra, and errors, are listed in Table~\ref{Table2}.

\begin{deluxetable}{ccccc} 
\centering 
\tablecaption{Observed Eclipse Spectra for CoRoT-2b.  Values are in parts-per-million.} 
\tablehead{\colhead{Wavelength (microns)} & \colhead{$\alpha$ Spectrum } & \colhead{Error} 
   & \colhead{$\beta$ Spectrum} &  \colhead{Error} }
\startdata 
  1.125 &  334.6 &   67.4 &  248.6 &   86.0 \\
  1.169 &  272.4  &  83.7  &  366.7  &  109.6 \\
  1.218 &  339.4 &  119.3 &  309.0  &  83.2 \\
  1.278 &  344.2 &   72.0 &  313.5  &  60.5 \\
  1.324 &  338.9  &  64.7 &  279.9  &  56.8 \\
  1.369 &  403.9  &  77.1 &  376.2  &  60.1 \\
  1.424 &  454.5  &  59.8 &  480.9  &  65.5 \\
  1.475 &  320.3  &  93.5 &  304.8  &  80.3 \\
  1.525 &  438.3  &  62.6 &  454.6  &  63.4 \\
  1.574 &  548.7  &  61.3 &  632.1  &  61.9 \\
  1.619 &  382.2  &  82.2 &  414.0  &  73.8 \\
\enddata 
\label{Table2} 
\end{deluxetable}

\section{Implications for the Atmosphere of CoRoT-2b}\label{sec:models}

No single model for the atmosphere of the planet fits all of the
available data to within the errors.  The observed properties of the
planet's atmosphere include: 1) the optical eclipse observed by the
~\emph{CoRoT} mission \citep{alonso09, snellen10}, 2) a ground-based
eclipse near 2\,$\mu$m \citep{alonso10}, 3) the overall level, general
slope with wavelength, and lack of obvious or known spectral features seen in our WFC3
data, and 4) eclipses in 3 Spitzer bands \citep{gillon10,
  drakespitzer}.  Figure~\ref{Fig14} shows these data in comparison to
several modeled spectra: a best-fit blackbody, conventional
solar abundance models \citep{burrows01, burrows08a, burrows08b,
  burrows10}, and a carbon-rich model \citep{madhu09, madhu10,
  madhu12}.  Although none of these are ideal fits to the data, each
model has characteristics that account for some observed properties of
the planet, as we now discuss.

\subsection{A Blackbody Spectrum?}

\begin{figure*}[t!] 
\epsscale{1.1} 
\plotone{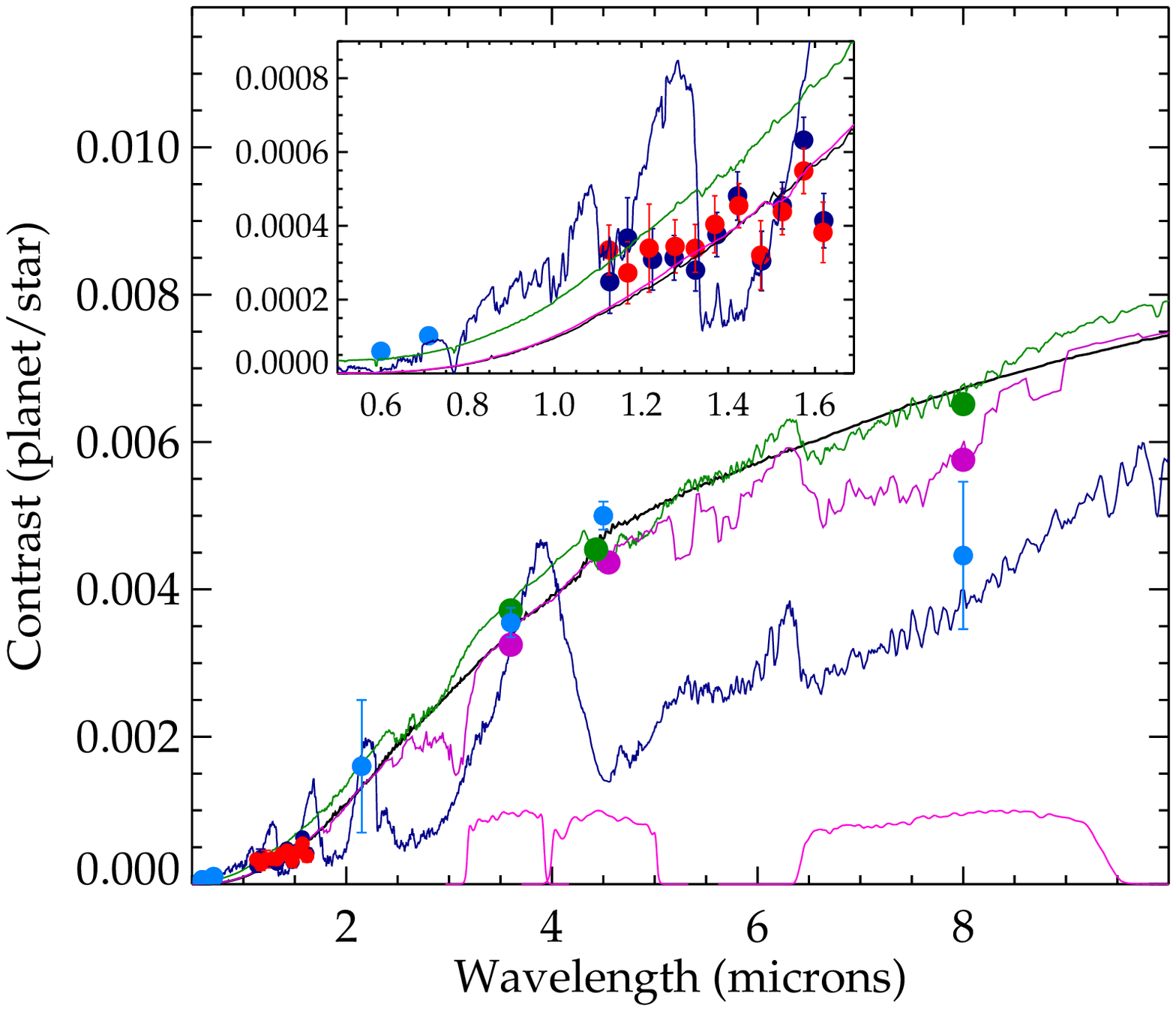}
\caption{Our WFC3 results for CoRoT-2b shown in the context of
  ground-based 2\,$\mu$m results \citep{alonso10}, the Spitzer results
  from \citet{drakespitzer}, and the optical eclipse depths from
  \citet{alonso09}.  The black line is an 1788K blackbody for the
  planet, and the dark blue line is a solar abundance clear atmosphere
  Burrows model previously used to interpret the Spitzer data
  \citep{drakespitzer}. The green line is the solar abundace Burrows
  model with additional continuous opacity (see text).  The magenta
  model is from Madhusudhan and has equal carbon and oxygen
  abundances.  All of the models lack temperature inversions (see
  text).  The inset shows our WFC3 results, from both our $\alpha$
  (red points) and $\beta$ (blue) analyses.  Note that the error in
  the overall level of the WFC3 points (Sec.~5.2) is much greater than
  the relative errors on individual points.}
\label{Fig14} 
\end{figure*}

The lower panel of Figure~\ref{Fig13} includes the contrast produced
by a best-fit blackbody for the planet compared to the results from
our $\alpha$ and $\beta$ analyses, and Figure~\ref{Fig14} plots that
blackbody in comparison to the totality of existing eclipse data.  We
adopt a Kurucz model for the star (Teff=5750, log(g)=4.5), yielding a
best-fit blackbody temperature of $1788\pm18$K for the planet in our
WFC3 band, from our $\alpha$-analysis.  This blackbody temperature
gives acceptable agreement with the infrared eclipse results at longer
wavelength (Figure~\ref{Fig14}).  The 1788K blackbody - derived from
the WFC3 data alone - misses the ground+Spitzer eclipse amplitudes by
an average of about $1.8\sigma$.  However, a blackbody spectrum for
the planet does not produce the best {\it slope} over the WFC3 band,
as we now discuss.

Our observed WFC3 spectrum for CoRoT-2b has two striking features: 1)
it slopes slightly upward with increasing wavelength, and 2) it shows
little to no evidence for water absorption or emission in the
1.4\,$\mu$m band. Statistically, the first question to resolve is
whether the simplest possible fitting function can account for our
spectrum.  The simplest function is a single value in contrast, i.e. a
flat line at the average contrast level. For our $\alpha$ analysis
spectrum (red points on Figure~\ref{Fig13}) the $\chi^2$ of the
best-fit flat line is 12.8 for 10 degrees of freedom, so our $\alpha$
analysis accepts a flat line as representing the planet's contrast
across the WFC3 band.  For our $\beta$ analysis (blue points on
Figure~\ref{Fig12}), the flat line $\chi^2$ is 28.6, rejecting the flat line at
$>99\%$ confidence.  So our $\beta$ analysis indicates a stronger and
more significant upward slope than does our $\alpha$ analysis.  That
is the single largest difference between our $\alpha$ and $\beta$
analyses, that are otherwise very consistent, with all points
overlapping within their error bars (Figure~\ref{Fig13}).  Both
of our WFC3 analyses reject the best-fit blackbody slope for the
planet, but only at about the 93\% confidence level.  The $\chi^2$
values are 17.0 and 17.8 (10 degrees of freedom) for our $\alpha$ and
$\beta$ spectra, respectively.  On the other hand, the blackbody is
obviously consistent with the weakness of water absorption in the WFC3
band.

We checked that our results are not affected by inadequate corrections
for detector non-linearity at the high fluence levels of our data.  We
repeated the $\beta$ analysis omitting the last (fourth) sample of the
exposure, and using only the first 3 samples, where the fluence level
($\sim$\,47,000 electrons) is well within the linear regime.  That
modified version of the $\beta$ analysis shows little difference from
the $\beta$ spectrum shown on Figs.~\ref{Fig13}~\&~\ref{Fig14} (but,
with larger errors due to the lower fluence levels).

The slope of the planet's spectrum across the WFC3 band is relevant to the
interpretation of the eclipse amplitude observed in the optical by CoRoT
\citep{alonso09, snellen10}.  If a 1788K blackbody agreed with the slope of our
observed spectrum, it would be reasonable to extrapolate that blackbody to judge the
magnitude of the thermal emission from the planet at optical wavelengths.  A
blackbody of 1788K would produce negligible thermal emission in the optical, and we
would conclude that the optical eclipses are due to reflected light.  However,
given that the observed slope across the WFC3 band does not decline as strongly
as a 1788K blackbody, it remains possible that the optical eclipses are due to
thermal emission.  That could happen, for example, if temperatures on the
star-facing hemisphere of the planet were spatially inhomogenous.  Hotter
regions having a small filling factor, combined with cooler regions of larger
filling factor, could in principle produce the observed slope across the WFC3
band, and account for the optical eclipses, while still remaining consistent
with the observed contrast at wavelengths exceeding 2\,$\mu$m.

In order to probe the viability of our speculation concerning temperature
inhomogeneities, we performed exploratory fits (not illustrated) using two
different blackbody temperatures and filling factors on the star-facing
hemisphere of the planet.  We find a good fit to our WFC3 and the CoRoT data
using $T_1 = 1500$K and $T_2 = 3600$K, with filling factors of $0.96$ and
$0.04$, respectively.  This combination matches the level of the contrast in the
CoRoT bands as well as the contrast level and wavelength dependence of our WFC3
results, but it significantly underestimates the contrast in the Spitzer bands
(by about 0.001).  Recent hydrodynamic simulations of hot Jupiter atmospheres
show brightness temperature variations as large as a factor of two on the
star-facing hemisphere of HD189733b \citep{dobbsdixon13}. Since that planet is
less strongly irradiated than CoRoT-2, the temperatures found by our exploratory
fits seem plausible.  Nevertheless, we do not here attempt to model the
atmosphere of CoRoT-2 using a self-consistent 3-dimensional approach
(temperature varying with depth and with horizontal coordinate).   Higher
quality data, such as we anticipate from the James Webb Space Telescope, may
justify such an approach in the future.

\subsection{Limit on WFC3 Spectral Features}

Both our $\alpha$ and $\beta$ spectra agree that a straight line
(contrast increasing linearly with wavelength) gives a good account of
our results across the WFC3 band: the $\chi^2$ values for a linear fit
(9 degrees of freedom) are 6.1 and 13.7 for our $\alpha$ and $\beta$
spectra, respectively.  These values leave little room for absorption
or emission by water vapor at 1.4\,$\mu$m. In order to specify a limit on the degree of water absorption or emission, we scale and fit a Burrows model to the data, using the model shown in blue on Figure~\ref{Fig14}. In order to make the limit responsive to the modulation caused by the actual water absorption (as opposed to the slope of the continuum), we allow for a linear baseline difference as a function of wavelength. We construct 10,000 trial data sets, adopting the error at each binned wavelength from our $\beta$-analysis, and we fit the model plus a linear baseline to each trial data set using linear regression. Based on the distribution of fitted amplitudes, we find an 85 ppm 3$\sigma$ limit on the amplitude of water absorption or emission, measured at the bandhead at 1.38\,$\mu$m. This limit assumes that the shape of the water absorption is the same as in the Burrows model. The $3\sigma$ limit of 85~ppm is significantly less than the
already weak water absorption seen during transmission spectroscopy of
the giant planets XO-1b and HD\,209458b \citep{deming13}, WASP-19
\citep{huitson, mandell}, and HAT-P-1b \citep{wakeford}. This
conclusion is significant, as can been seen by reference to one
conventional solar abundance Burrows model \citep{burrows01,
  burrows08a, burrows08b, burrows10} illustrated as the dark blue line
on Figure~\ref{Fig14}.  This model is not intended as a fit to the WFC3 data,
but it was invoked by \citet{drakespitzer} in an attempt to account
for the Spitzer observations.  Although it misses the 4.5\,$\mu$m
Spitzer point, \citet{drakespitzer} discussed the possibility of
circumplanetary carbon monoxide emission in that band, due to tidal
stripping by the star.  However, this model produces a much larger
spectral modulation in the WFC3 band than is seen in our observed
spectra.

\subsection{Solar Abundance Model Atmospheres}

CoRoT-2b is an unusual planet, and the Spitzer data have been
particularly difficult to understand, as discussed by
\citet{drakespitzer} (but, see \citealp{madhu12}).  The relatively
high contrast at 3.6- and 4.5\,$\mu$m seems to require a hot
continuum, allowing little if any molecular (principally water)
absorption. Simultaneously, the lower contrast at 8\,$\mu$m requires
absorption to a significant degree. We here explore the potential for
conventional solar abundance model planetary atmospheres to account for
the totality of the CoRoT-2b eclipse data.

The weakness of absorption features can be produced in a solar
abundance model by adding continuous opacity by small particle
scattering and/or absorption.  That could dampen features in the
emergent spectrum at short wavelengths, but a reduced scattering
cross-section with increasing wavelength could allow greater spectral
contrast at 8\,$\mu$m (mentioned by \citealp{drakespitzer}).  If the
temperature remains nearly constant as a function of pressure/altitude
in the planet's atmosphere, that would also suppress any absorption or
emission features in the emergent spectrum.  Figure~\ref{Fig14} shows
the contrast from a Burrows model \citep{burrows01, burrows08a,
  burrows10} having three additional sources of opacity not present in
a clear atmosphere.  This model is shown in green on
Figure~\ref{Fig14}, and has redistribution parameter $P_n=0.1$
\citep{burrows08a}.  The additional opacity sources are first, a high
altitude optical (0.4-1.0\,$\mu$m) absorber of opacity
$0.2\,cm^2g^{-1}$.  Second, an absorbing haze opacity of
$0.04\,cm^2g^{-1}$ uniformly distributed at all pressures and
wavelengths, and third, a scattering opacity of $0.08\,cm^2g^{-1}$
also uniformly distributed at all pressures and wavelengths. The
scattering opacity acts to increase the reflected light, but not
increasing the thermal emission. Note that in principle we could
include a wavelength dependence to the opacity of the broadly
distributed hazes, but we prefer to keep this {\it ad hoc} opacity as
simple as possible.

The uniformly distributed hazes damp the spectral modulation in the
WFC3 bandpass to an acceptable degree, but the model misses the
overall WFC3 contrast level, being too high by 161 ppm.  Like all
single-spatial-component models, it's slope across the WFC3 band is
larger than our data. Given the error level of our white light eclipse
($395^{+69}_{-45}$\,ppm), the overall contrast difference is
significant at $2.3\sigma$, which is the single largest problem with
this model.  On the other hand, the scattering opacity increases the
contrast in the optical to the point where it underestimates
the~\emph{CoRoT} eclipse amplitude by less than $2\sigma$.  Also,
among the models we've tested, it does the best job of reproducing the
long wavelength eclipse amplitudes ($1.5\sigma$ on average).

The aggregate eclipse data are ambiguous concerning the possibility of
a thermal inversion in the dayside atmosphere of CoRoT-2b.  As
discussed in \citet{madhu12}, the lower brightness temperature in the
8\,$\mu$m Spitzer bandpass compared to the brightness temperatures in
the shorter wavelength channels (except 4.5\,$\mu$m) suggests a
temperature profile decreasing outward in the atmosphere.  If that
gradient is flatter than radiative equilibrium models predict, it will
help to account for the lack of strong spectral features.  On the
other hand, the solar abundance radiative equilibrium model (green
line on Figure~\ref{Fig14}) achieves good agreement with the
4.5\,$\mu$m Spitzer eclipse depth by incorporating $0.2\,cm^2g^{-1}$
of extra optical-wavelength opacity at low pressures ($\sim$\,1~mbar)
close to where radiation in the 4.5\,$\mu$m band is formed
\citep{burrows07}.  Indeed, that model shows a temperature rise of
about 75\,Kelvins, near 0.2 mbars.  But due to the more widely
distributed absorbing haze, the temperatures in this model at high
altitude are already hundreds of Kelvins over the values they would
have in a clear atmosphere.  To the extent that this model is
preferred, or that a flattened temperature gradient counts as a
weakly-inverted atmosphere, then CoRoT-2b could be claimed to have a
temperature inversion. However, this evidence for an inversion is
weaker than for HD\,209458b \citep{burrows07, knutson08}, and is
ambiguous in the sense that the atmosphere could be heated without
satisfying a strict definition of inversion (temperature increasing
with height).  \citet{knutson10} hypothesized that planets orbiting
active stars will not have strong atmospheric temperature inversions,
because the absorbing species that causes the inversion (e.g.,
\citealp{Hubeny:03, Fortney:08}) may be destroyed by the enhanced UV
flux from stellar activity. CoRoT-2a is an active star
\citep{spotted}, and lack of a strong thermal inversion in CoRoT-2b
would support the \citet{knutson10} hypothesis.

\subsection{A Carbon-rich Model Atmosphere}

An alternate way to reduce the spectral modulation by water vapor in
the WFC3 bandpass is to reduce the equilibrium water vapor mixing
ratio, for example by increasing the carbon abundance relative to
oxygen.  This also helps to decrease absorption in the 3.6- and
4.5\,$\mu$m bands (although methane does contribute some absorption at
3.6-\,$\mu$m), while preserving absorption at 8\,$\mu$m via the
7.8\,$\mu$m methane band.  We used the methodology described by
\citet{madhu09} and \citet{madhu10} to find a possible carbon-rich
match to the aggregate data for this planet (except for the optical
eclipses). \citet{madhu12} discussed CoRoT-2b and was able to fit the
pre-WFC3 data by varying the C/O ratio to various degrees.  The
magenta line on Figure~\ref{Fig14} is a model with an enhanced carbon
abundance (C/O=1), and having a non-inverted atmosphere with modest
thermal contrast (700K increase in temperature from upper boundary to
the optically thick photosphere). The enhanced carbon abundance
weakens the water absorption, but allows sufficient absorption near
8\,$\mu$m to account for that Spitzer point to within $\sim\,1\sigma$.
The average agreement with the ground+Spitzer eclipses is $1.9\sigma$,
not quite as good as the blackbody and the solar abundance model.  On the
other hand, the carbon-rich model does the best job -- of the atmospheric models, i.e., beyond just a linear fit -- of reproducing the
WFC3 spectrum ($\chi^2=16.1$ for 10 DOF), and in particular it agrees
essentially perfectly with the amplitude of the WFC3 white-light
eclipse.  It does not require additional haze opacity.

\subsection{Reprise of the Model Atmosphere Comparisons}

We here summarize the main conclusions from comparing the aggregate
eclipse data for this planet with emergent spectra from different
models.  We tested a blackbody, as well as more sophisticated solar
abundance and carbon-rich models.  No model fits all of the data.  The
limit on spectral modulation due to water absorption in the WFC3 band
is our main observational result.  Given the lack of clear and
unequivocal molecular absorption features in the WFC3 and other bands,
emergent spectra more sophisticated than a blackbody are unproven. A
blackbody spectrum gives an acceptable fit to the data except for the
optical eclipses as seen by \emph{CoRoT}.  A blackbody spectrum fits
the slope over the WFC3 band poorly, but multi-component blackbodies
due to spatial inhomgeneities on the star-facing hemisphere have the
potential to help account for the observations, including the optical
eclipses as seen by \emph{CoRoT}, especially if extra scattering
opacity increases toward short wavelengths.  Note that the absorbing
and scattering hazes invoked in our solar abundance model are
qualitatively similar to extra absorption and scattering opacity
inferred for the archetype planet HD\,189733b \citep{pont,evans}.

Although a blackbody spectrum reasonably accounts for the infrared
eclipse data, it is not a model of the planet's atmosphere {\it per
  se}.  Instead, the planetary atmosphere can mimic a blackbody via
the presence of extra continuous opacity that damps the observed
thermal contrast, or due to a high carbon abundance that weakens the
bands of the principal molecular absorber (water vapor).  In either
case, extra scattering opacity at optical wavelengths could help to
account for the amplitude of the optical eclipses.  We find only weak
evidence for a strong temperature inversion, but extra absorbing opacity in
the solar abundance model would perturb the temperature profile in a
manner similar to a temperature inversion, but less extreme.

\section{Summary} 

We observed the Very Hot Jupiter CoRoT-2b in secondary eclipse using
three visits by the WFC3 G141 grism on HST. Even without utilizing the
new spatial scan mode, we obtained spectra with errors approaching the
photon noise limit. We characterized the instrument-related systematic
effects present in the data. We find a time-dependent variation in the
background intensity, a visit-long slope, slopes associated with each
orbit, and we investigate the `hook' effect that occurs after data
transfers. We explored the behavior, dependencies, and how best to
account for these effects in data analyses. In particular, we defined
the amplitude of the hook effect versus the exposure level in
electrons (Figure~6).

We measure the thermal emission from the planet in the 1.1-1.7\,$\mu$m
band, but we find no spectral features to a $3\sigma$ limit of 85
ppm. We used a differential method to derive the spectrum and cancel
the systematic errors \citep{deming13}, obtaining results close to
photon-limited.  No model fits all available eclipse data for this
planet to within the errors.  We consider solar abundance and
carbon-rich spectral models, as well as a simple blackbody spectrum,
to account for the eclipse data.  The spectral models do not clearly
surpass the blackbody spectrum in terms of the quality of the fit.
The slope of the data within the WFC3 bandpass is less than given by
all of the models, including the best-fit blackbody. There is weak and
ambiguous evidence that the atmospheric temperature structure is
inverted, but a reduced temperature gradient may be present, and may
help to mimic the quasi-blackbody nature of the emergent spectrum.
Extra atmospheric continuous opacity is a strong possibility to
account for the lack of spectral features in the WFC3 band.  If that
opacity has a scattering component, it can help to account for the
optical eclipse amplitude of this planet as observed by CoRoT.
Spatial inhomogeneities in temperature on the star-facing hemisphere
may also help to account for the optical eclipse and the slope of the
spectrum in the WFC3 band.

 \bibliographystyle{apj} 
 
\end{document}